\documentclass[aps,prd,twocolumn,nofootinbib,showpacs,amssymb,amsmath,floatfix]{revtex4}

\usepackage{graphicx}
\usepackage{epsfig}
\usepackage{subfigure}

\def \aaps{Astron.\ Astrophys.\ Supp.\ }
\def \aap{Astron.\ Astrophys.\ }
\def \apjl{Astrophys.\ J.\ }

\def \apj{Astrophys.\ J.\ }

\def \mnras{Mon.\ Not.\ Roy.\ Astron.\ Soc.\ }

\def \nat{Nature\ }

\renewcommand{\d}{\textrm{d}}
\newcommand{\der}[2]{\frac{{\d}#1}{{\d}#2}}   
\newcommand{\mchi}[0]{m_\chi c^2}

\newcommand{\sigva}[0]{\left\langle \sigma_{\text{A}}v\right\rangle}

\def\alt{\raise0.3ex\hbox{$\;<$\kern-0.75em\raise-1.1ex\hbox{$\sim\;$}}}
\def\agt{\raise0.3ex\hbox{$\;>$\kern-0.75em\raise-1.1ex\hbox{$\sim\;$}}}

\newcommand{\bw}{\begin{widetext}}
\newcommand{\ew}{\end{widetext}}
\def\d{{\rm d}}

\begin{document}

%

\title{Radio and gamma-ray constraints on dark matter annihilation
in the Galactic center} 

\author{
  R.~M.~Crocker$^{1,2,3}$\footnote{roland.crocker@mpi-hd.mpg.de},
  N.~F.~Bell$^1$\footnote{n.bell@unimelb.edu.au},
  C.~Bal\'azs$^{2}$\footnote{csaba.balazs@sci.monash.edu.au}, and
  D.~I.~Jones$^{3}$\footnote{djones@mpi-hd.mpg.de}
}

\affiliation{
 $^1$School of Physics, University of Melbourne, 3010, Victoria, Australia \\
 $^2$School of Physics, Monash University, 3800, Victoria, Australia \\
 $^3$Max Planck Institut f{\" u}r Kernphysik, Saupfercheckweg 1, D-69117 Heidelberg, Germany
}

\date{\today}

\begin{abstract}
We determine upper limits on the dark matter (DM) self-annihilation cross
section for scenarios in which annihilation leads to the production of
electron--positron pairs.  In the Galactic centre (GC), relativistic
electrons and positrons produce a radio flux via synchroton emission, and
a gamma ray flux via bremsstrahlung and inverse Compton scattering.
On the basis of archival, interferometric and single-dish radio data,
we have determined the radio spectrum of an elliptical region around
the Galactic centre of extent 3$^\circ$ semi-major axis (along the
Galactic plane) and 1$^\circ$ semi-minor axis and a second,
rectangular region, also centered on the GC, of extent $1.6^\circ
\times 0.6^\circ$.  The radio spectra of both regions are non-thermal
over the range of frequencies for which we have data: 74 MHz --
10 GHz. We also consider gamma-ray data covering the same region from
the EGRET instrument (about GeV) and from HESS (around TeV).
We show how the combination of these data can be used to place robust
constraints on DM annihilation scenarios, in a way which is relatively
insensitive to assumptions about the magnetic field amplitude in this
region.  Our results are approximately an order of magnitude more
constraining than existing Galactic centre radio and gamma ray limits.
For a DM mass of $m_{\chi}=10$\,GeV, and an NFW profile, we find
$\sigva \leq \textrm{few} \times 10^{-25}$\,cm$^3$s$^{-1}$.
\end{abstract}


\maketitle

\section{Introduction}

It is a remarkable fact that the identity of most of the matter in the
Universe is unknown.  An abundance of observational evidence allows us
to infer the existence of dark
matter~\cite{Kamionkowski_review,Bertone_review,Bergstrom_review} via
its gravitational influence.  However we have as yet no direct
detection and very little information about its corpuscular
properties.  In this paper, we focus on one of the fundamental
particle properties of dark matter: its self annihilation cross
section.

If the dark matter were in thermal equilibrium in the early Universe,
the annihilation cross section would determine the relic density in
the Universe today.  A velocity averaged annihilation cross section of
$\langle \sigma_A v \rangle_{\textrm{th}} \sim {\rm few} \times
10^{-26}~\textrm{cm}^3 \textrm{s}^{-1}$ is required in order to
produce the relic abundance of $\Omega_{\textrm{DM}} \sim 0.2$.
However, it is not necessary for DM to be a thermal relic, in which
case both (significantly) larger or
smaller cross sections are possible.
Regardless of the thermal history, $\langle \sigma_A v \rangle$ 
controls the annihilation
rate in DM halos in the Universe today, thus determining the size of
detectable signals emanating from regions of DM concentration.  It is
these annihilation fluxes that potentially permit the technique of
indirect detection of dark matter.

Recent cosmic ray positron/electron data from a number of experiments
have led to much excitement about indirect DM detection.  Anomalies in
the cosmic ray positron spectra have been reported by the PAMELA,
Fermi and HESS experiments, implying an apparent excess of positrons
beyond those due to conventional astrophysical processes.
%
PAMELA~\cite{pamela} has observed a rise in the $e^+/(e^- + e^+)$ flux
at energies above approximately 10 GeV, while recent data from Fermi
LAT~\cite{fermi} and HESS~\cite{hess} show an excess in the $(e^- +
e^+)$ flux up to and beyond 1 TeV, respectively.  (ATIC~\cite{atic}
and PPB-BETS~\cite{ppbbets} observed a similar excess, however, with
considerably higher uncertainty than Fermi.)

The explanation of these positron excesses is far from clear, and, in
light of these, researchers have been motivated to examine or
re-examine conventional cosmic ray interaction~\cite{Stawarz2009} and
propagation models~\cite{Cowsik2009,Katz2009}, acceleration of
$e^+e^-$ in cosmic ray sources~\cite{blasi, Hu:2009bc, Dado:2009ux},
electron and positron emission from supernova
remnants~\cite{Shaviv:2009bu, Fujita:2009wk}, and the production of
positrons by pulsars~\cite{hooper, yuksel_kistler, profumo,
  Malyshev:2009tw, Barger:2009yt, Grasso:2009ma, Mertsch:2009ph,
  Malyshev:2009zh}.  As an alternative to conventional astrophysical
mechanisms, it has also been speculated that DM annihilation or decay
in the Milky Way may be the source of the excess electrons and
positrons.

Many authors have proposed models in which electron and positron
fluxes arise from DM annihilation~\cite{Cirelli:2008pk,
  ArkaniHamed:2008qn, Cholis:2008qq, Harnik:2008uu, Allahverdi:2008jm,
  Calmet:2009uz, Shirai:2009fq, Chen:2009mj, Hamaguchi:2009jb,
  Okada:2009bz, Fukuoka:2009cu, Bai:2009ka, Shirai:2009wi,
  Chen:2009zp, Mardon:2009gw, Demir:2009kc, Hooper:2009gm,
  Choi:2009qc, Feldman:2009wv} or decay~\cite{Yin:2008bs,
  Hamaguchi:2008rv, Ibarra:2008jk, Nardi:2008ix, Ishiwata:2008qy,
  DeLopeAmigo:2009dc, Arvanitaki:2009yb, Buchmuller:2009xv,
  Ibarra:2009dr, Chen:2009gd}, either directly or indirectly.  (See
also, Ref.~\cite{He:2009ra} and references therein.)  Note that, in
contrast to the positron data, the PAMELA antiproton/proton
observations do not indicate an anomalous
contribution~\cite{pamela_antiproton}.\footnote{For completeness, we
  mention that both the reality of an anomaly in the PAMELA
  $e^+/(e^-+e^+)$ flux and the absence of such in the anti-proton flux
  have been questioned~\cite{Katz:2009yd,Kane:2009if}.}  Therefore, DM
models which feature significant hadronic annihilation modes are
constrained, while leptonic channels are preferred.
However, in order to account for the observed positron spectra,
$\langle \sigma_A v \rangle$ must be significantly enhanced above the
expected thermal relic value.  Such an enhancement could be of
astrophysical origin, e.g., a boost due to significant substructure
throughout the halo or a local clump of dark
matter~\cite{Hooper:2008kv, Brun:2009aj}.  Alternatively, it could be
of particle physics origin, e.g., a Breit-Wigner resonant
enhancement~\cite{Feldman:2008xs,Ibe:2008ye, Guo:2009aj,Bi:2009uj} or
the Sommerfeld effect (see, e.g.,
Refs.~\cite{Hisano:2004ds,Cirelli:2008pk, ArkaniHamed:2008qn,
  MarchRussell:2008yu}) in which low velocity annihilation (i.e. in
Galactic halos) is enhanced while early Universe freeze-out is
affected to a lesser
degree~\cite{Dent:2009bv,Zavala:2009mi,Feng:2009hw,Backovic:2009rw}.

Several techniques may be used to constrain the production of $e^+$
and $e^-$ within Galactic halos, all of which rely on other
accompanying observational signals.
Charged particle production in a halo is necessarily accompanied by
photon signals, including gamma rays, X-rays, microwaves and radio
waves.  These signals are produced via the various energy loss
processes that charged particles undergo, examples of which include
synchrotron emission in Galactic magnetic fields, inverse Compton
scattering of electrons from interstellar radiation field, and
bremsstrahlung.
Charged particle production is also accompanied by {\it internal}
bremsstrahlung radiation~\cite{ib1,ib2,ib3,ib4,bell_jacques}, which is
an electromagnetic radiative correction, and is {\it not} due to
interaction in a medium.

In this work, we shall use the electron energy loss processes to derive 
constraints on $\langle \sigma_A v \rangle$.  Annihilation channels we 
consider include both direct production of monoenergetic $e^\pm$, and 
channels in which a spectrum of secondary $e^\pm$ are generated via decays 
of primary annihilation products, such as $\overline{q}q$.
Note that, given their identical distributions and radiative
signatures (at the energies under consideration), we shall imply both
electrons and positron when we refer to ``electrons" in this work,  
unless explicitly noted otherwise.
Our analysis is distinguished from previous work in this area by the
use of a new synthesis~\cite{Crocker2009} of existing Galactic center
radio data which allows us to derive particularly sensitive
constraints.  In addition, we make a careful study of the interplay of
various energy loss processes, and show that a combination of
different techniques leave our final bounds relatively insensitive to
uncertainties in magnetic field amplitude.

A number of astrophysical uncertainties enter our calculations.  As
with all indirect detection analyses, we have a sensitivity to the
assumed dark matter halo density profile.  In addition we are subject
to uncertainties in the Galactic magnetic field amplitude, the
background light density, and the gas density.
Variation in the assumed magnitudes of these parameters alters the
proportion of electron cooling that takes place via the various energy
loss processes. Higher magnetic field strengths, for example, lead to
greater synchrotron losses, while higher background light density
leads to more inverse Compton scattering.  However, while the
astrophysical assumptions control the {\it relative} importance of the
various electron energy loss processes, the eventual {\it total}
energy loss is a constant. (Note that, as we show below, relativistic
electrons in the GC lose their energy -- via whatever exact
combination of processes -- before they are transported out of the
region.)  Therefore, while constraints from a single process (say,
synchrotron radiation) are individually quite uncertain, we show that
the combined constraints derived using all energy loss processes are
robust and feature only mild dependence on these astrophysical
parameters.

This paper is structured as follows: in section \ref{AstrInput} we
discuss the radio and $\gamma$-ray data which provide the empirical
constraints on the various DM models we test.  We also describe the
physical environment at the GC (magnetic field, ambient hydrogen
number density) which determines how the cooling and subsequent
radiation from relativistic electrons proceeds and the DM profiles we
investigate.  Section \ref{DMsignals} describes the distribution of
electrons putatively {\it injected} by DM annihilations and explains
how this injection spectrum is shaped into a steady state distribution
by the various cooling processes.  We also set out here our calculation of the
radio and $\gamma$-ray emission by these steady state electron
distributions. In section \ref{DMCnstrnts} we compare predicted
emission from DM annihilation (parameterized in terms of DM
particle mass and velocity averaged cross section) against empirical
data and delineate the regions of parameter space we can thereby
exclude. We also compare our results with bounds existing in the
literature. Section \ref{Conclusion} contains our concluding remarks.

\section{Astrophysical Inputs}\label{AstrInput}

\subsection{Radio Data}

\begin{figure}
\vspace{1pc}
\includegraphics[width=\columnwidth]{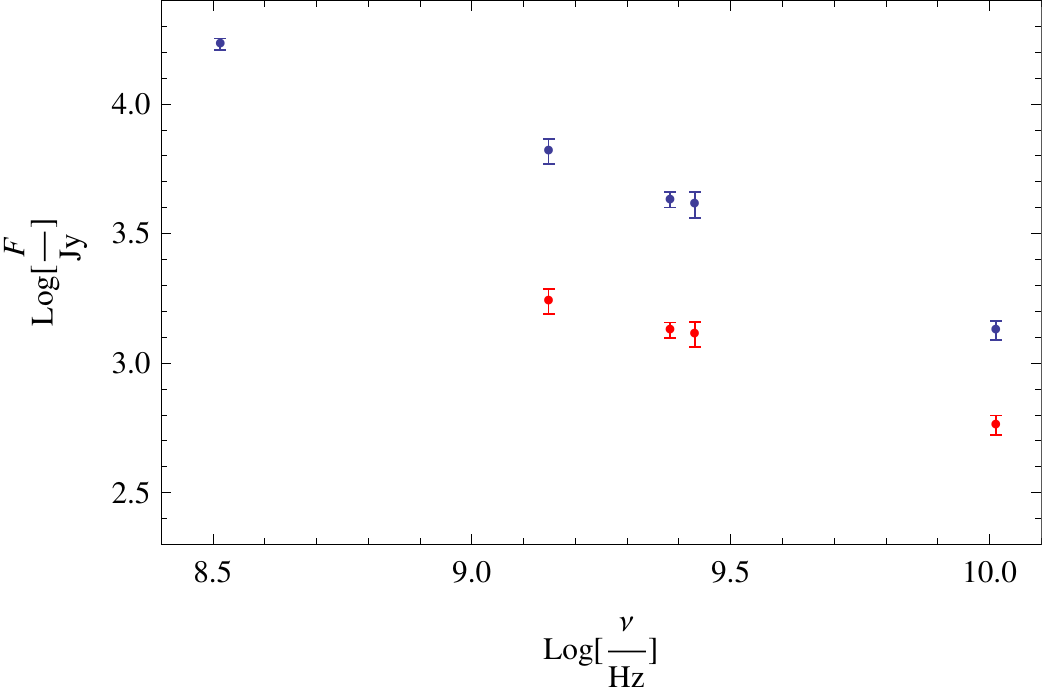}
        \caption{Radio spectra of the DNS (blue, upper) and HESS (red, lower) regions.\label{fig:DNSRadioSpectrum}}
        \label{f1}
\end{figure}

LaRosa {\it et al.}~\cite{LaRosa2005} observed a discrete but diffuse,
non-thermal radio source ({\bf DNS}) 
covering a roughly elliptical 
region around the Galactic centre (GC) of extent 3$^\circ$ semi-major
axis (along the Galactic plane) and 1$^\circ$ semi-minor axis between
74 and 330 MHz. This angular region corresponds, more-or-less, to the
usual definition of the Galactic nuclear bulge~\cite{Ferriere2007}.
Subsequently, the work of Crocker {\it et al.}~\cite{Crocker2009}
assembled archival, interferometric and single-dish radio data that
demonstrates that the DNS radio structure is evident at frequencies up
to 10 GHz and has a non-thermal radio spectrum over the full 74 MHz --
10 GHz range of the data (see Fig.~\ref{fig:DNSRadioSpectrum} for the
spectrum and refer to Table~\ref{table:radioData} for references and
other pertinent information on each radio datum). It should be noted
though that, significantly for the purposes of Crocker {\it et al.},
the spectrum of the DNS can {\it not} be described as a pure power
law, but rather exhibits a spectral down-break at about a GHz.  For a
full account of the radio data processing and analysis supporting this
conclusion the reader should refer to~\cite{Crocker2009}, but we
repeat a number of particularly relevant points below.

We also consider the constraints that arise from radio spectrum of the
smaller region defined by $| l | < 0.8^\circ$ and $| b | < 0.3^\circ$.
We refer to this as the HESS region, as it
corresponds to a region for which the gamma ray intensity is reported
by the HESS collaboration.

\begin{table}[h]
	\centering	
	 \begin{tabular}{|c|c|c|c|c|c|}
	 \hline\hline
         $\nu$ (GHz) & Telescope &  Beam & Flux density & Error & Ref. \\
	\hline
	.074$^*$ & VLA &  2$'$ & 16,200 Jy&1,000 Jy & \cite{LaRosa2005} \\
	.330$^\dag$ & Green Bank  & 39 & 18,000 Jy& 5\% & \cite{LaRosa2005} \\
	1.408 & Effelsberg  &  9.4$'$ & 7,300 Jy& 10\% & \cite{Reich1990} \\
	2.417 & Parkes  & 10.4$'$ & 4,900 Jy& 6\% & \cite{Duncan1995} \\
	2.695 & Effelsberg  &  4.3$'$ & 4,400 Jy& 10\% & \cite{Reich1984} \\
	10.29 & Nobeyama  & 2.9$'$ & 1,400 Jy& 7\% & \cite{Handa1987} \\
	\hline
        \end{tabular}

       	       	\caption{Surveys and Observational data used to derive the spectrum for the $6^\circ\times2^\circ$ region centred on the GC. Notes: $^*$At 74 MHz the large-angle  Galactic plane synchrotron background/foreground flux contribution is not measured (and hence not accounted for) due to the interferometric nature of the VLA observation. $^\dag$Total flux supplied by Dr Crystal Brogan (private communication).}
	\label{table:radioData}
\end{table}

Radio data at 74 MHz~\cite{LaRosa2005} were obtained by the Very Large Array (i.e., an interferometer), while data at all higher frequencies come from single dish instruments. (Because of the problem of free-free absorption along the line-of-sight to the GC~\cite{LaRosa2005} we do not employ the 74 MHz datum as a constraint in our fitting below. At higher frequencies, because of the $\propto \nu^{-2.1}$ dependence of the free-free optical depth, absorption by this process is not a significant factor.)
The radio flux at every frequency under consideration has or might have contributions from i) synchrotron emission by cosmic ray electrons along the line of sight, i.e., both in front of and behind the GC along the Galactic plane; ii) synchrotron emission by  cosmic ray electrons gyrating in the GC itself; iii) discrete, astrophysical sources in the GC and along the line-of-sight (these include optically thin and optically thick thermal bremsstrahlung emitters and those with non-thermal spectra); iv)  diffuse, large scale HII
regions (i.e., regions of ionized atomic hydrogen that produce thermal bremsstrahlung); and v) (possibly and as investigated in this paper) relativistic electrons and positrons injected following annihilation or decay of putative DM particles.

As detailed in Ref.~\cite{Crocker2009}, the contribution of discrete sources (taken to be emission on angular scales $\lesssim 1.2^\circ$) can be measured in the single dish data by Fourier analysis. (At 330 MHz the contribution of discrete sources was measured directly with the VLA~\cite{LaRosa2005}). This discrete source contribution is at a $<20$\% level at all frequencies under consideration and has been removed from every radio datum we use in our analysis. Otherwise, we remove no other astrophysical contribution and, in considering possible limits on synchrotron emission by relativistic electrons and positrons injected by DM annihilation or decay, we take the conservative approach~\cite{Bertone2009} of requiring that predicted synchrotron emission be no larger than 3$\sigma$ above any radio datum.

\subsection{Gamma-Ray Data}

In addition to synchrotron radiation, bremsstrahlung and
(sub-dominantly) inverse Compton (IC) emission will inescapably be
generated by any relativistic electron population given the presence
of background gas (mostly molecular hydrogen) and the GC's background
light field.  
Electrons with GeV scale energies produce both bremsstrahlung photons
of approximately 100 MeV and synchrotron radiation at GHz frequencies
(given the range of magnetic field amplitude that is physically
plausible for the GC; see below).
This means that $\gamma$-ray observations by the EGRET instrument
(onboard the Compton Gamma-Ray Observatory) are particularly
complementary to the radio data introduced above, as both signals are
produced by electrons of the {\it same} energy.
Finally, additional $\gamma$-ray photons are also generated from
neutral meson decay in any scenario involving quark-pair production in
DM annihilation and both $\bar{q}q$ and $e^+e^-$ scenarios generate internal bremsstrahlung $\gamma$-rays.

\subsubsection{EGRET Data}

We claim a conservative upper limit to the integral gamma ray
intensity from the DNS region. This is obtained from Figs. 2(c) and
2(d) of Hunter et al.~\cite{Hunter1997} which, when combined, show a
pedestal in the super-300 MeV intensity of $\sim 3 \times 10^{-4}$
cm$^{-2}$ s$^{-1}$ sr$^{-1}$, averaging over $|b| < 2^\circ$ and
within the longitude defined by $|l| < 30^\circ$. Given that the DNS
region is much smaller in extent (in longitude) than the pedestal,
and that there are no structures evident in these longitude-dependent
intensity plots on the scale of the DNS, we estimate an upper-limit to
any exotic contribution to the intensity in the DNS region at the
level of twice the largest actual excursion in the intensity above the
pedestal value, viz. $1 \times 10^{-4}$ cm$^{-2}$ $s^{-1}$
sr$^{-1}$. We are confident that a more detailed analysis would
produce a more stringent constraint. In any case, results pertaining
to the region from the Fermi (GLAST) telescope are eagerly awaited.

\subsubsection{HESS Data}

Similarly, bremsstrahlung and IC emission by the higher energy members
of the electron population will generate $\sim$TeV gamma rays (as will internal bremsstrahlung and, where applicable, neutral meson decay).  
The
{\it differential} intensity in this energy range may be compared with the
(conservative) upper limit obtained from observations by
HESS~\cite{Aharonian2006}.  This instrument has detected diffuse
emission over the approximate energy range 300 GeV -- 10 TeV and {\it over the smaller solid angle} defined by $| l | < 0.8^\circ$
and $| b | < 0.3^\circ$. The TeV intensity recorded is $1.4 \times 10^{-20}$ cm$^{-2}$ eV$^{-1}$
s$^{-1}$ sr$^{-1}$  with only limited and dimmer diffuse
emission detected outside this region but within the DNS field. 
The HESS data analysis does, however, include a background subtraction 
from a nearby region, and this must be accounted for in our analysis\citep{bell_jacques,Mack2008}.
Because of this background subtraction,  in the case of an NFW profile, a DM scenario may predict an absolute flux of $\gamma$-rays at TeV energies and within the HESS field up to 2 $\times$ the
observed $\sim$TeV intensity.
For the isothermal profile, on the other hand, because of the very flat distribution and the consequently very small relative difference in $\gamma$-ray intensity predicted for the HESS region and the background region, no competitive constraint is proffered by the $\sim$TeV data at all (and we, therefore, do not plot TeV $\gamma$-ray constraint curve for this profile).

\subsection{Ambient Environment}

\subsubsection{Gas Density}
\label{section_GasDensity}

Over the DNS region, the volumetric-average hydrogen number density can be calculated to be about 13 cm$^{-3}$ on the basis of the data presented in Ref.~\cite{Ferriere2007}\footnote{We assume the DNS volume to be an elliptical spheroid with circular cross-section in the Galactic plane (i.e., its greatest line-of-sight extent is equal to its width along the plane). This implies a total volume of $\sim 3.0 \times 10^{63}$ cm$^3$.}. Much of this density can be ascribed to molecular hydrogen found at very high number density ($> 10^{3.5}$ cm$^{-3}$) but small filling factor ($<0.01$) in the cores of the unusually-dense, Galactic center molecular cloud population~\cite{Paglione1998,Ferriere2007}. Quite possibly electrons impinging from outside the dense molecular material may be excluded from it~\cite{Gabici2007,Crocker2007,Protheroe2008}.  Excluding this very high density phase ($n_H > 10^3$ cm$^{-3}$) but including hydrogen in relatively low density molecular hydrogen, atomic hydrogen (HI) and `warm' ionized hydrogen (H$^+$) one finds a minimum (path-integrated) $n_H$ for a DNS electron of $\sim$ 2.7 cm$^{-3}$ and $\sim$ 23 cm$^{-3}$. We adopt these values in our calculations henceforth. (The putative `very hot', plasma phase of the DNS, that would have a filling factor of $\sim 85$\%, would have lower number density than this. However, relativistic electrons would not be trapped in this phase~\cite{Higdon2007} and, in any case, the existence of this phases has recently been thrown into considerable doubt: see below.)

\subsubsection{Galactic Magnetic Field}

The amplitude of the GC magnetic field remains uncertain by about two orders of magnitude.
\begin{itemize}
\item On the basis of the 74 and 330 MHz observations of the DNS (nuclear bulge) introduced above, LaRosa et al.~\cite{LaRosa2005} use the equipartition argument to suggest that the field on DNS size scales is only around 6 $\mu$G -- very similar in amplitude to that typical for the Galactic disk -- climbing to only about 10 $\mu$G over the inner $1.^\circ6$ along the Galactic plane. (Coincidentally, the region over which the HESS collaboration measures a diffuse $\sim$ TeV $\gamma$-ray intensity~\cite{Aharonian2006}.) These smaller field amplitudes have recently received support from rotation measures to external galaxies close to the GC~\cite{Roy2008}. 

\item A rather stronger field over the DNS region is suggested by the work of Spergel and Blitz~\cite{Spergel1992} who note that the putative ``very hot" (8 keV) X-ray emitting plasma found throughout the GC~\cite{Koyama1989}  and the unusually dense and turbulent molecular gas might be in pressure equilibrium (implying near equipartition between these ISM phases). In analogy with conditions local to the solar system~\cite{Webber1998}, it might then be expected that the magnetic field also contribute a roughly equal pressure requiring it to have about 100 $\mu$G amplitude. The existence of the very hot plasma has recently been cast into extreme doubt, however, with the apparently diffuse X-ray emission being, it is contested~\cite{Revnivtsev2009}, essentially ascribable to unresolved, point X-ray sources. This would essentially nullify the Spergel and Blitz argument. 

\item Finally, (coincidentally or not) over the HESS field, radio observations of the `non thermal filaments' (synchrotron-emitting radio structures, unique to the inner Galaxy, that run essentially perpendicular to the Galactic plane) suggest that the ambient field might be as strong as 1 mG~\cite{Yusef-Zadeh1987,Morris1989,Morris2007}. 
\end{itemize}
The situation regarding the GC magnetic field is entirely unclear, therefore. 

Recently this argument has been joined by Crocker et
al.~\cite{Crocker2009}.  On the basis of a simultaneous analysis of
the higher-frequency part of the radio spectrum of the DNS radio
structure and $\sim$ GeV $\gamma$-ray data covering the same region
(i.e., the very data described above), these authors have recently
claimed a probable value for the magnetic field in this region of 100
$\mu$G with a lower limit (at 95\% confidence) at 50 $\mu$G.  
However, we cannot self-consistently assume this result holds for the
purposes of the current paper.  The analysis in Crocker et al. assumes
that the observed radio emission from the DNS region can be attributed
to ``conventional" astrophysical electrons governed by a pure power
law at injection, and that any spectral features detected in the
observed spectrum arise in the cooling of the injection spectrum
(which is initially featureless).
If we allow for the possibility that DM makes an additional
contribution to the electron injection spectrum, 
the kinematics of DM annihilation allows for the
possibility of
spectral features in the {\it injection} spectrum of the {\it in situ} electron population, 
so the assumption of Crocker et al.  that cooling alone introduces spectral features does not necessarily hold.
(E.g., one can find a fit acceptable at better than 2$\sigma$ confidence to the radio spectrum invoking synchrotron emission by a population of relativistic electrons injected following the annihilation of dark matter particles with a mass $\sim$ 30 GeV and assuming the quark hadronization spectrum given below. We certainly do not claim this as a detection!)
This implies we cannot self-consistently adopt the
magnetic field strength inferred in Ref.~\cite{Crocker2009} in this work.

Given the uncertainties explained above, in this paper we investigate
DM constraints across the whole range of magnetic field amplitudes
supported by the existing literature, which we take to be 10 -- 100
$\mu$G for the DNS region and 10 -- 1000 $\mu$G for the HESS diffuse
flux region defined above.  However, as we demonstrate below, radio
and $\gamma$-ray data offer complimentary constraints over this range:
radio data provide the more severe restriction on $\langle \sigma_A
v\rangle$ for strong fields, while $\gamma$-ray data are more
constraining for weak fields (because, in the latter case, there is
less synchrotron suppression of the high-energy electrons required for
IC up-scattering of ambient light into $\sim$GeV and $\sim$TeV energy
ranges).

\subsection{Dark matter distribution}\label{DMD}

We employ the usual Navarro-Frank-White (NFW)
distribution~\cite{Navarro:1996gj}
\begin{equation}
    \label{NFW}
    \rho(r)=\frac{\rho_{\text{h}}}{\frac{r}{r_{\text{h}}}\left(1+\frac{r}{r_{\text{h}}}\right)^2}
    \,\,\, ,
\end{equation}
where $\rho_{\text{h}} = 0.572$ GeV cm$^{-3}$ and $r_{\text{h}}$ = 14
kpc.  The above implies a volumetric-average DM density of $3.6 \times
10^{10}$ eV cm$^{-3}$ over the DNS region and $1.2 \times 10^{11}$ eV
cm$^{-3}$ over the HESS region. We employ these volume-average
quantities in our calculations (given we assume a one-zone model).

We also investigate limits in the case that the DM distribution follows a truncated isothermal profile. 
Here we assume the parametrization presented in Ref.~\cite{Bertone2009} following Ref.~\cite{Bahcall1980}:
\begin{equation}
    \label{ISO}
    \rho(r)=\frac{\rho_{\text{s}}}{1+\left(\frac{r}{r_{\text{s}}}\right)^2}
    \,\,\, ,
\end{equation}
where $\rho_{\text{s}} = 1.16$ GeV cm$^{-3}$ and $r_{\text{s}}$ = 5
kpc.

\section{Dark matter signals}\label{DMsignals}

\subsection{Dark matter annihilation and primary electron spectrum}


In typical models with weakly-interacting massive particles (WIMPs), 
the DM masses fall in the GeV-TeV range,
while the DM annihilation cross section is given approximately by the
thermal relic value of $\langle \sigma_A v \rangle_{\textrm{th}} \sim {\rm few}
\times 10^{-26} \textrm{cm}^3\textrm{s}^{-1}$.  However, the DM self
annihilation cross section can be much smaller if coannihilations play
a significant role in determining the WIMP freezeout density, as
occurs in many supersymmetric DM scenarios.  Conversely, the
Sommerfeld effect can boost the self annihilation cross section
relevant for galactic halos in the Universe today, effectively
resulting in values much larger than the thermal relic expectation.
Moreover, if dark matter was populated via some non-thermal mechanism,
$\langle \sigma_A v \rangle_{\textrm{th}}$ is not relevant to the presently 
observed average abundance.  It is
thus appropriate to take $\langle \sigma_A v \rangle$ to be a free
parameter, spanning a wide range of possible values.  Likewise, we
shall also take the DM mass to be a free parameter.  (Indeed, masses
above 1 TeV have been considered in light of recent cosmic ray
positron results and data from Fermi LAT.)
We shall consider two DM annihilation processes: $1.~\chi\chi
\rightarrow \bar{q}q$ and $2.~\chi\chi \rightarrow e^+e^-$.

\subsubsection{$\chi\chi \rightarrow \bar{q}q$}

Annihilation to $\bar{q}q$ is taken as a representative prototype of
many DM models, as the resulting electron energy spectrum is a good
approximation to that for many annihilation channels.  For example,
the annihilation spectrum for a specific supersymmetric WIMP candidate
may be calculated with the DarkSUSY package~\cite{Gondolo:2004sc}.
However, the annihilation channels $\chi\chi\rightarrow \bar{q}q, ZZ,
W^+W^-$ all lead to essentially the same electron energy spectrum
(while the electron energy spectrum resulting from
$\chi\chi\rightarrow \tau^+\tau^-$ is significantly different).
Note that for annihilation to $ZZ$, $W^+W^-$, and $\bar{q}q$,
hadronization or decay of the primary annihilation products leads, via
charged pions and muon intermediaries, to positrons and electrons with
a broad spectrum of energies.

For simplicity, we shall assume full annihilation into $\bar{q}q$. For
this channel, $e^-$ and ($e^+$) will be produced by decaying muons
(anti-muons) produced in charged pion decay.  The resulting $e^+$ {\bf
  +} $e^-$ spectrum can be written, following Borriello et
al.~\cite{Borriello2009}, in a simple polynomial form of the ratio
$E_e/\mchi$:
\begin{equation}
    \label{dNdE}
    \der{N_e}{E_e}(E_e)=\frac{2}{\mchi}\sum_{j\in
    J}^{}a_j\left(\frac{E_e}{\mchi}\right)^j \, ,
\end{equation}
where $J=\{-\frac{3}{2},-\frac{1}{2},0,\frac{1}{2},2,3\}$ and the
coefficients $a_j$ are listed in Table \ref{tab:ajvalues} taken from~\cite{Borriello2009}.

This analytical expression is a reasonable approximation for $E_e \agt
1$ GeV.  At low energies the analytical expression from~\cite{Borriello2009} has an asymptotic behaviour $\propto E_e^{-1.5}$
while the kinematics of the charged meson and muon decay chains mean
that the actual spectrum has a low-energy cut-off. 
Because of the large magnetic fields we sample, our calculated radio
flux from synchrotron emission is potentially sensitive to this effect
at the lowest frequencies considered (cf.~\cite{Borriello2009}).
We therefore explicitly introduce a cut-off by patching on a numerical
description of the low-energy spectrum we have previously
obtained~\cite{Crocker2007} in the context of a numerical treatment of
secondary electron production from hadronic cosmic ray interactions
(where identical kinematical considerations come into play) to
Borriello et al.'s distribution.

\begin{table}[ht]
    \centering
    \caption{$a_j$ values from \cite{Borriello2009}}
    \vspace{1mm}
\begin{tabular}{|c|c|}
\hline
\textbf{coefficient}  & \textbf{numerical} \\
\hline
$a_ {-3/2}$                                                                & $0.456$       \\
$a_ {-1/2}$                                                                     & $-5.37$       \\
$a_ {0}$                        & $10.9$        \\
$a_ {1/2}$                                                                                                      &   $-6.77$     \\
$a_ {2}$     &   $0.969$     \\
$a_ {3}$      &   $-0.185$    \\
\hline
\end{tabular}
    \label{tab:ajvalues}
\end{table}

\subsubsection{$\chi\chi \rightarrow e^+e^-$}

We shall also consider direct annihilation to monoenergetic electrons,
for which the electron energy spectrum per annihilation is simply
given by
\begin{equation}
    \der{N_e}{E_e}(E_e)= 2 \delta(m_\chi-E_e).
\end{equation}

Note that the injection electron spectra differ from the steady state
electron spectra due to the action of electron cooling, as discussed
below.

\subsection{Steady-state electron distribution}

The injection spectrum of electrons produced in the Galaxy, at position $r$, as a result of Dark Matter annihilation is
\begin{equation} \label{inj1}
Q(E_e,r)=\frac{1}{2}\left(\frac{\rho(r)}{m_{\chi}}\right)^2 \sigva
\der{N_e}{E_e}\ .
\end{equation}
In general, the injected electrons loose energy interacting
with the interstellar medium and move away from the production
site. Both these broad processes lead to modifications of the shape of the injection spectrum. 
If, however, the transport timescale is much longer than the cooling timescale for electrons in the relevant energy range one is in the
so-called ``thick-target" regime in which case the steady-state (cooled) electron distribution can be written as
  \begin{equation}
\frac{d n_e}{d E_e}(E_e, \vec{r}) = 
{\int_{E_e}^ {m_\chi c^2} \!\!\!\!\! \,\,\,dE_{e}' \,\,Q(E_{e}',\vec{r}) \over - dE_e(E_e)/dt} 
\label{steady}
\end{equation}
 \noindent
where $dE_e(E_e)/dt$ is the cooling rate, resulting from
the sum of several energy loss processes that affect electrons (described immediately below).
We show in the Appendix that we are justified in adopting the thick target regime for the GC environment.

\subsection{Electron cooling processes}

\begin{figure}
\includegraphics[width=\columnwidth]{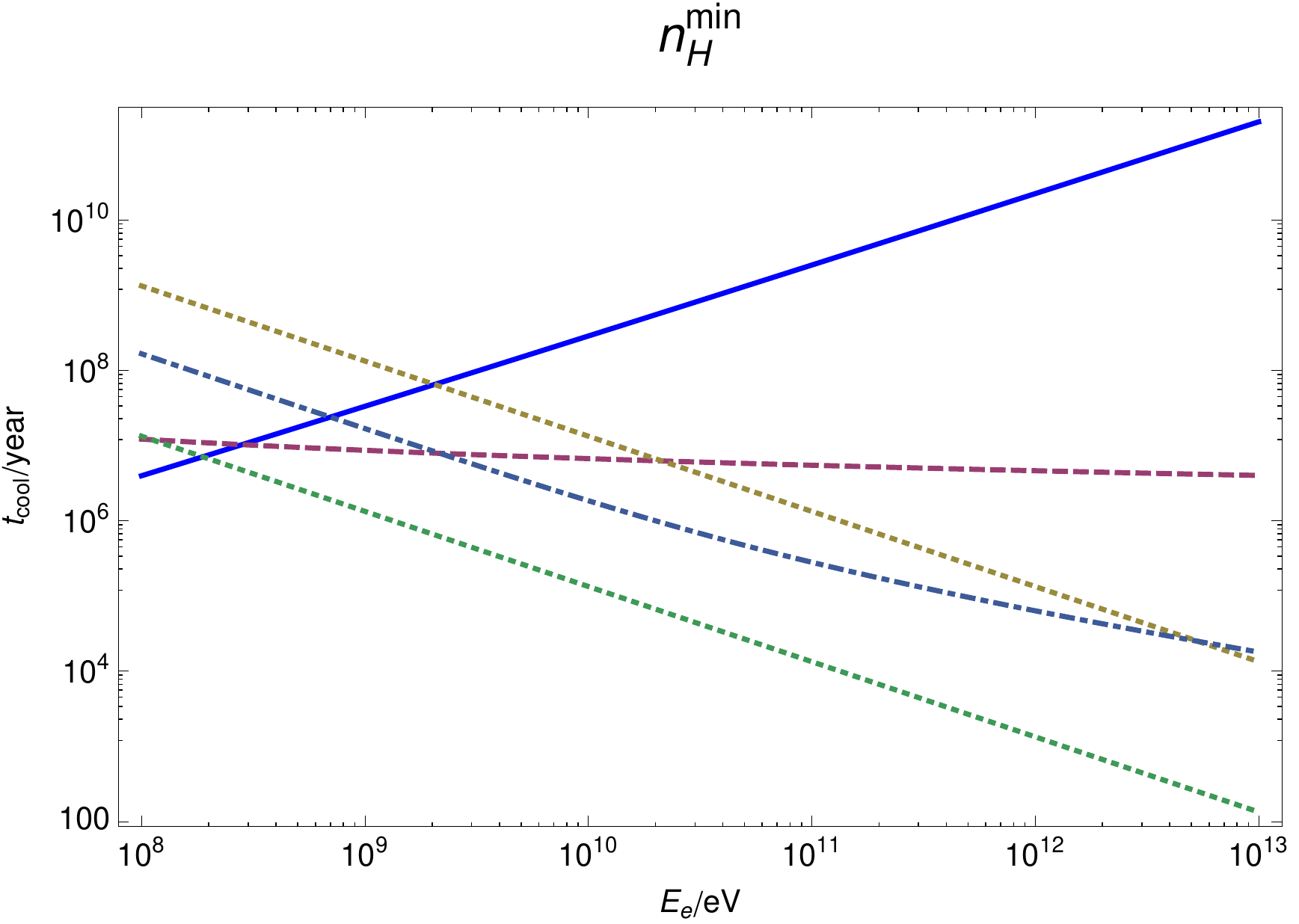}
\caption{Cooling time-scale due to various processes affecting electrons and positrons in the GC environment as a function of  energy. Curves are -- {\bf solid}: ionization, {\bf dashed}: bremsstrahlung; {\bf dotted}: synchrotron for (upper) 10 $\mu$G and (lower) 100 $\mu$G; and {\bf dot-dashed}: inverse Compton.}
\label{f3}
\end{figure}

\begin{figure}
\includegraphics[width=\columnwidth]{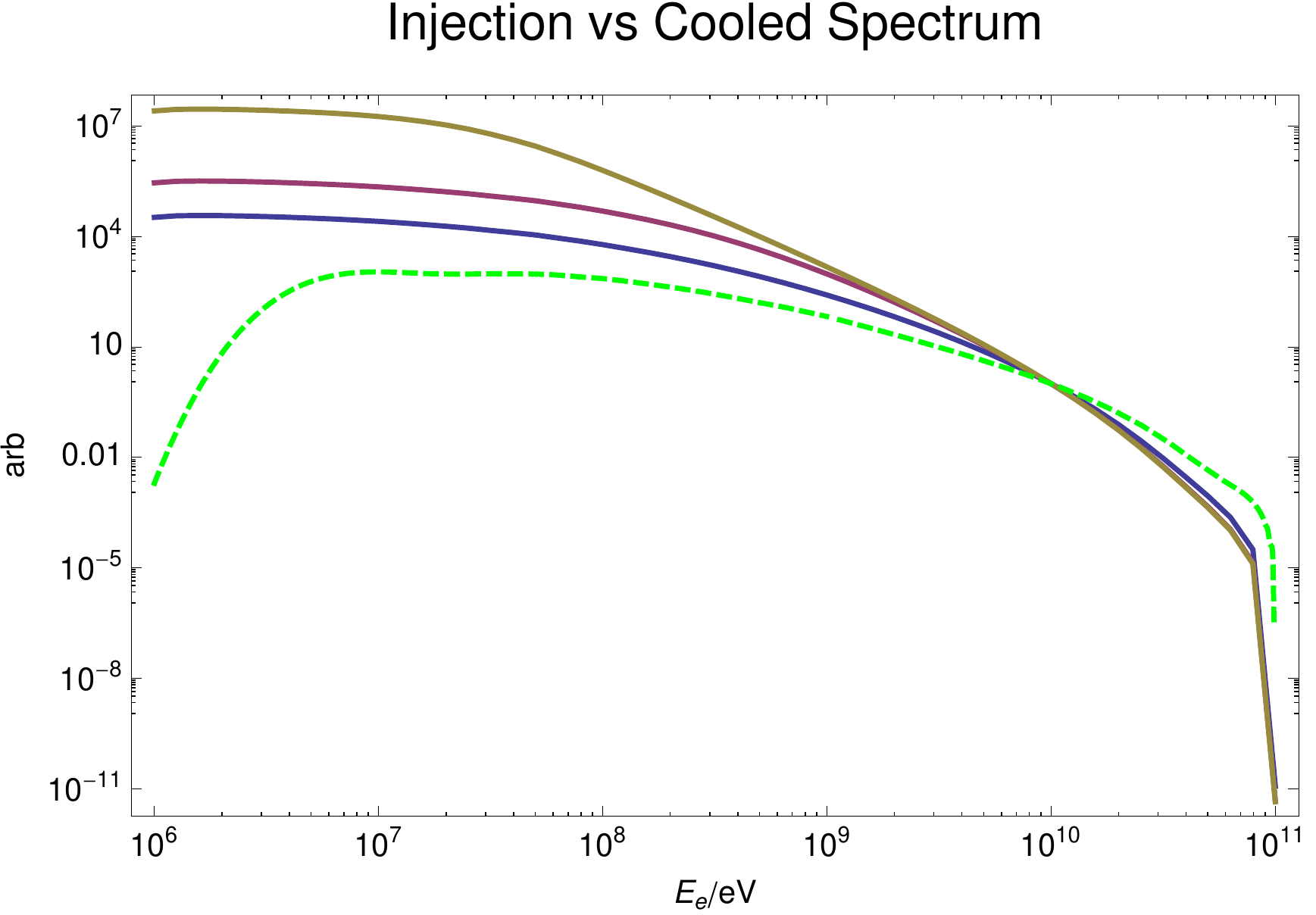}
\caption{{\it Shape}  of the injected electron + positron distribution (green; dashed) and steady state distributions (for various magnetic field amplitudes -- mG (yellow; solid, upper); 100 $\mu$G (purple; solid, center); and 10 $\mu$G (blue; solid, lower). The curves have been normalized to 1 (on an arbitrary scale) at an energy of 10 GeV. Note the low-energy, kinematic cut-off in the injected distribution. The plot assumes $m_\chi = 100$ GeV.}
\label{plotSpctrm}
\end{figure}

In the GC environment four cooling processes are potentially important for charged leptons with energies such that they radiate at radio and $\gamma$-ray wavelengths (electrons and positrons suffer essentially identical energy losses over the relevant energy scale and are treated identically). These processes are dominantly (from low to high energy) ionization, bremsstrahlung, and synchrotron and/or IC emission. We show the cooling timescales, as a function of electron (or positron) energy, in Fig.~\ref{f3} for plausible GC environment parameters. In Fig.~\ref{plotSpctrm} we show examples of the {\it shape} of the injected electron (and positron) spectrum together with the steady-state electron distribution after cooling.

\subsection{Radiative Processes}

Of the cooling processes listed above, all except ionization lead to
potentially-observable signatures at radio and $\gamma$-ray
wavelengths. The amount of cooling that takes places via either radio
or gamma ray emission -- and the constraints derived on the basis
of that channel -- depend on assumptions made about the
astrophysical inputs.  We calculate separate constraints from radio
and gamma ray emission, in each case choosing astrophysical inputs
conservatively. In particular, for the hydrogen density, $n_H$, we
conservatively choose a value which leads to smallest observational
signal for each process considered.  The magnetic field strength is
left as a free parameter to be varied, and we shall demonstrate the
our final results are relatively insensitive to this parameter.
In contrast, we fix the ambient hydrogen number density to be
the appropriate minimal value set out in \S \ref{section_GasDensity} (given the consideration that most of the ambient gas is `locked-up' in molecular cloud cores of very high density but extremely low filling factor).
We describe our treatment of each radiative process below.
Representative $\gamma$-ray spectra from each of the relevant process (for a $m_\chi = 1$ TeV DM mass and $\sigva = 10^{-26}$ cm$^{3}$ s$^{-1}$, with a Borriello et al.~\cite{Borriello2009} $e^\pm$
  spectrum) are shown in fig.\ref{plotSampleSpectrum}.

\subsubsection{Synchrotron}

We use standard formulae~\cite{Rybicki1979} to calculate the radio emissivity of relativistic electrons (and positrons) in the Galactic center magnetic field. 

\subsubsection{Bremsstrahlung Calculation}

We use standard formulae (see Ref.~\cite{Longair1992}, p.~84 et seq.) to calculate the bremsstrahlung emissivity of relativistic electrons (and positrons) in the Galactic center environment.  

\subsubsection{Inverse Compton Calculation}

For inverse Compton cooling and emission calculations we assume the full interstellar radiation field determined for the inner 500 pc of the Galaxy in~\cite{Porter2006}. This has an energy density around 19 eV cm$^{-3}$. We calculate the Compton scattering emissivity of relativistic electrons using the full Klein-Nishina cross section set out in section 2.3.3 of Ref.~\cite{Sturner1997}. For the minimal $n_H$ value we assume, the integrated energy flux (100 MeV+) from IC emission is about 70\% as large as the bremsstrahlung energy flux.

\subsubsection{Gamma-rays from neutral meson decay}

\begin{figure}
\vspace{1pc}
\includegraphics[width=\columnwidth]{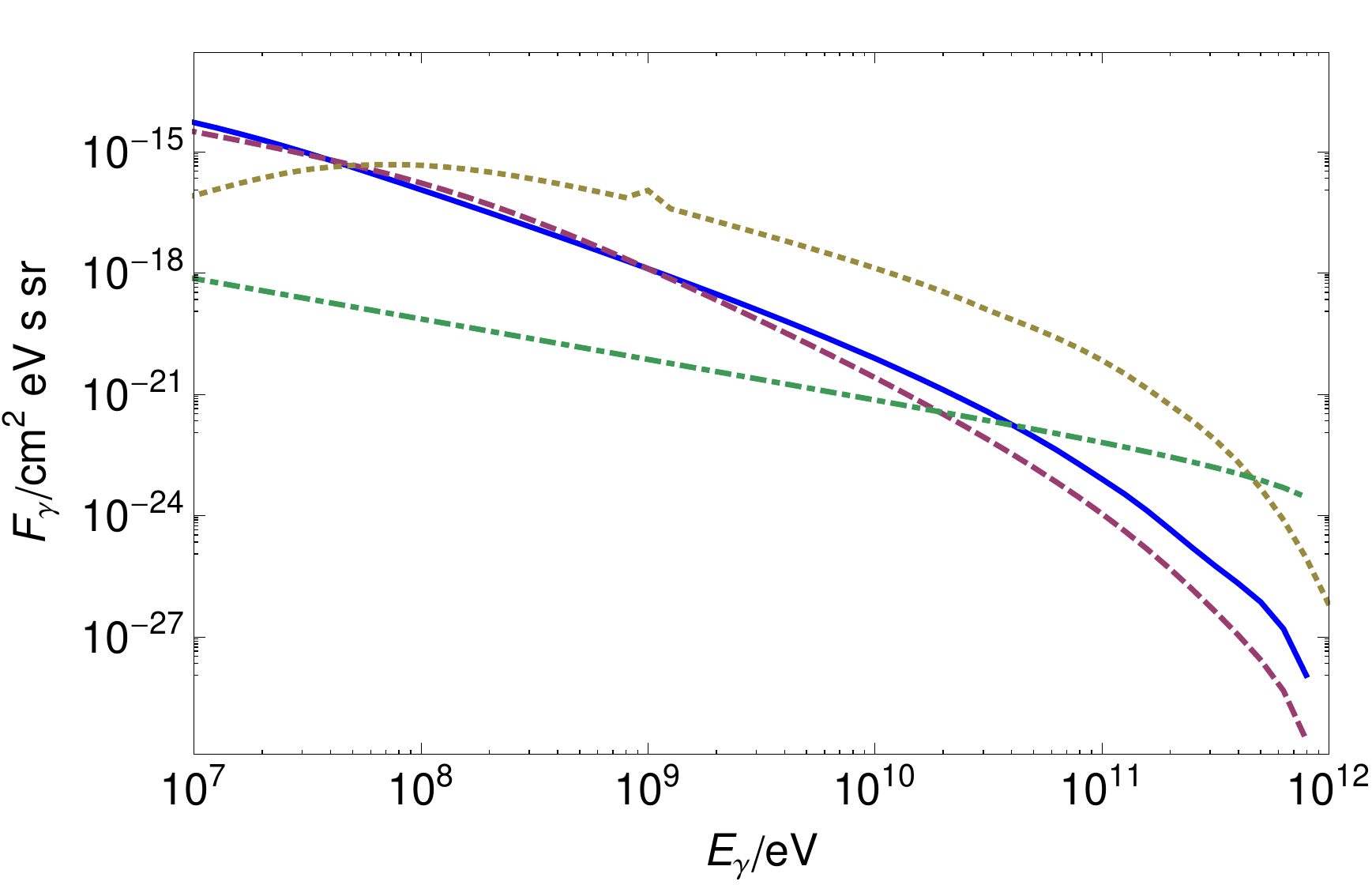}
        \caption{Sample $\gamma$-ray spectra of the DNS region for a $m_\chi = 1$ TeV DM mass and $\sigva = 10^{-26}$ cm$^{3}$ s$^{-1}$, with a Borriello et al.~\cite{Borriello2009} $e^\pm$
  spectrum, a plausible ambient field of 100 $\mu$G, and an ambient hydrogen number density of  $\sim 3$ cm$^{-3}$.
  Curves are: (blue, solid) inverse Compton; (purple,dashed) bremsstrahlung; (yellow,dotted) neutral meson decay (following hadronization); and (green, dot-dash) internal bremsstrahlung.}
        \label{plotSampleSpectrum}
\end{figure}

For DM annihilation into $q \bar{q}$ as investigated by Borriello et al., a 
$\gamma$-ray signal from the decay of  neutral mesons produced in
quark hadronization is inescapable. 
The contribution of this process is often dominant for relevant energies and environmental parameters (see fig. \ref{plotSampleSpectrum}). We employ a 
parameterization of the numerical results of Bergstr{\"o}m et al.\cite{Bergstrom1998} for the $\gamma$-ray spectra produced by quark hadronization following annihilation of 500 GeV neutralinos (displayed in the their figure 12 and parameterized in terms of the scaling variable $x \equiv E_\gamma/m_\chi$).
A representative distribution is that following $u\bar{u}$ production:
\begin{equation}
\frac{d N_\gamma}{d x} = \exp(-6.8 x) \ x^{-1.5}
\end{equation}
We also assume a branching ratio of 100\% into the maximally massive quark pair allowable given the DM mass (which means either $b\bar{b}$ or $t\bar{t}$ over the range of $m_\chi$ we explore).

\subsubsection{Gamma-rays from internal bremsstrahlung}

DM annihilation to any charged particles is necessarily accompanied by
{\it internal} bremsstrahlung (IB)
radiation~\cite{ib1,ib2,ib3,ib4,bell_jacques}.  This is an
electromagnetic radiative correction, and not to be confused with the
regular bremsstrahlung considered in this paper.  IB produces hard
gamma rays up to the dark matter mass, with an approximately
model-independent spectrum, largely free from both particle and
astrophysics uncertainties.  For the (sub-dominant) contribution of
internal bremsstrahlung we employ the parameterization of Refs.~\cite{ib2,bell_jacques}:
\begin{equation}
\frac{d N_\gamma}{d E_\gamma} = \frac{\alpha}{\pi E_\gamma} \left[\ln\left(\frac{s'}{m_e^2}\right) - 1\right]  \left[1 + \left(\frac{s'}{s}\right)^2 \right] \, ,
\end{equation}
where $s = 4m^2_\chi$ and $s' = 4m_\chi(m_\chi - E)$.
(This equation has to modified for the case of $q\bar{q}$ production by replacing the electron mass with the effective quark mass, and $\alpha$
with $Q_q^2 \ \alpha$, where $Q_q$ is the electric charge of quark $q$~\cite{ib3}.)

\section{DM Annihilation constraints}\label{DMCnstrnts}

\subsection{Discussion of results}

Figures \ref{qNFW}, \ref{eNFW}, \ref{qIso} and \ref{eIso}, display the
bounds we obtain for the annihilation cross section, as a function of
DM mass.  We show results corresponding to our four main scenarios,
namely, the $\chi\chi\rightarrow\bar{q}q$ and $\chi\chi\rightarrow
e^+e^-$ processes, for both NFW and isothermal DM halo profiles.  For
each scenario, we show how the constraints depend upon the assumed
magnetic field strength within the plausible range $10-100\mu$G.

Neglecting the $\sim$TeV $\gamma$-ray constraints for the moment, 
in each scenario investigated $\gamma$-ray data is the most
constraining at low magnetic field amplitude, and radio the best at
high.  Adjusting the magnetic field amplitude alters the proportion of
electron cooling that occurs via synchrotron emission.  Thus, for low
(high) $B$ amplitude the radio flux is less (more) constraining.
The total energy loss of the electron population is controlled by a
combination of the available cooling mechanisms, such that reducing
the amount of cooling taking place by synchrotron emission leads to
relatively more gamma rays emission via bremsstrahlung and inverse
Compton scattering.
(Note that any individual electron must eventually loose all its energy
via one mechanism or another, and it is impossible for this energy
loss to occur without the production of detectable signals.)
Therefore while the radio constraints, considered alone, are quite
sensitive to the assumed magnetic field strength, our final results,
being the strongest of either the radio or gamma constraints at a given
$B$ amplitude, are much less sensitive to the assumed value of $B$.

For the NFW profile, we find that the $\sim$TeV $\gamma$-ray
constraint invariably becomes the best for larger values of $m_\chi$.
(As remarked above, because of the background subtraction procedure,
TeV $\gamma$-ray data does not proffer a competitive constraint for
relatively flat profiles like the isothermal).
The $\sim$TeV $\gamma$-ray constraint curve is highly structured as a result of it being a comparison between predicted and observed differential $\gamma$-ray intensities at the nine separate energies the HESS collaboration has listed starting at $\sim$300 GeV  in concert with the three (for $e^+e^-$) of four (for $\bar{q} q$) $\gamma$-ray production mechanisms and their different
thresholding effects with respect to the  $m_\chi$ parameter.
A somewhat similar effect -- though less pronounced is see in the radio constraints. 
In contrast, the EGRET $\gamma$-ray constraints (taken above a single photon energy of 300 MeV in
the integral intensity) define a fairly smooth constraint for each
scenario whereas .

In Fig.~\ref{plotBestConstrntAllFields} we compare the best constraint
obtained (from any modality except for $\sim$TeV $\gamma$-rays) for
plausible magnetic field values, namely 10, 30, 100, 1000 $\mu$G.
We see that the best constraint on $\langle \sigma v \rangle$ varies
by a factor of $< 10$ as $B$ is varied from 10-100 $\mu$G. (This is to
be compared with a larger variation in the radio constraint alone of
$\sim 30$.)
Therefore, when considered in totality, the $\gamma$-ray and radio
data conspire to generate constraints on DM mass/annihilation cross
section that are roughly constant across the reasonable range of
astrophysical parameters (in particular, magnetic field amplitude).
For larger magnetic field amplitude (i.e. $B = 1000 \mu$G, which is at
the upper end of the astrophysically plausible range) synchrotron
emission dominates the cooling, and somewhat stronger constraints are
obtained.  This occurs because the field has become strong enough that
the GHz range synchrotron-radiation is sampling lower energy (and
therefore more numerous) electrons.

Comparing the figures for the NFW and isothermal profiles
(e.g. Fig.~\ref{qNFW} with Fig.~\ref{qIso}, or Fig.~\ref{eNFW} with
Fig.~\ref{eIso}) it is evident that the exclusion curves for the NFW
profile are always much more constraining than for the corresponding
isothermal scenario.  This is due to the high central DM density of
the strongly peaked NFW profile, in comparison with the low central
DM density of the flatter isothermal profile

Likewise, the radio observation of the HESS region (small solid angle
at the GC) provide more stringent constraints than the radio
observations of the DNS region (larger solid angle) for the peaked NFW
profile.
For the isothermal profile, however, we have something of the opposite situation:
the radio bounds for the DNS regions are often more constraining
than those for the HESS region.  This can be understood since, for the
flat isothermal profile, the average DM density is approximately the
same for both regions, while the radio background for the DNS region
is lower than for the HESS region.


With the exception of the HESS $\sim$TeV $\gamma$-ray constraint, the
$\bar{q}q$ injection scenario tends to generate somewhat tighter
constraints than the monoenergetic $e^+e^-$ spectrum, for all values
of $m_\chi$ larger than $\sim$ 100 GeV.  In the $\bar{q}q$ case,
electrons and positrons are injected as secondaries from meson decay
following quark hadronization.  Therefore, a single $\chi\chi
\rightarrow \bar{q} q$ interaction produces more electrons and
positrons (though of much lower energy) than for the monoenergetic
$\chi\chi \rightarrow e^+e^-$ case, and thus generally tighter limits.
In addition, the production of $\gamma$-rays from neutral meson decay,
leads to strong MeV gamma ray limits in the case of $\bar{q}q$.
At the lowest energies (below $\sim 100$ GeV), the $e^+e^-$ radio
limits surpass those for $\bar{q}q$, due to the higher average energy
of the injection spectrum.

\begin{widetext}

\begin{figure}
\subfigure[~$\chi\chi \rightarrow \bar{q}q$; NFW profile; $B= 10 \mu$G.]{
\includegraphics[width=0.31\columnwidth]{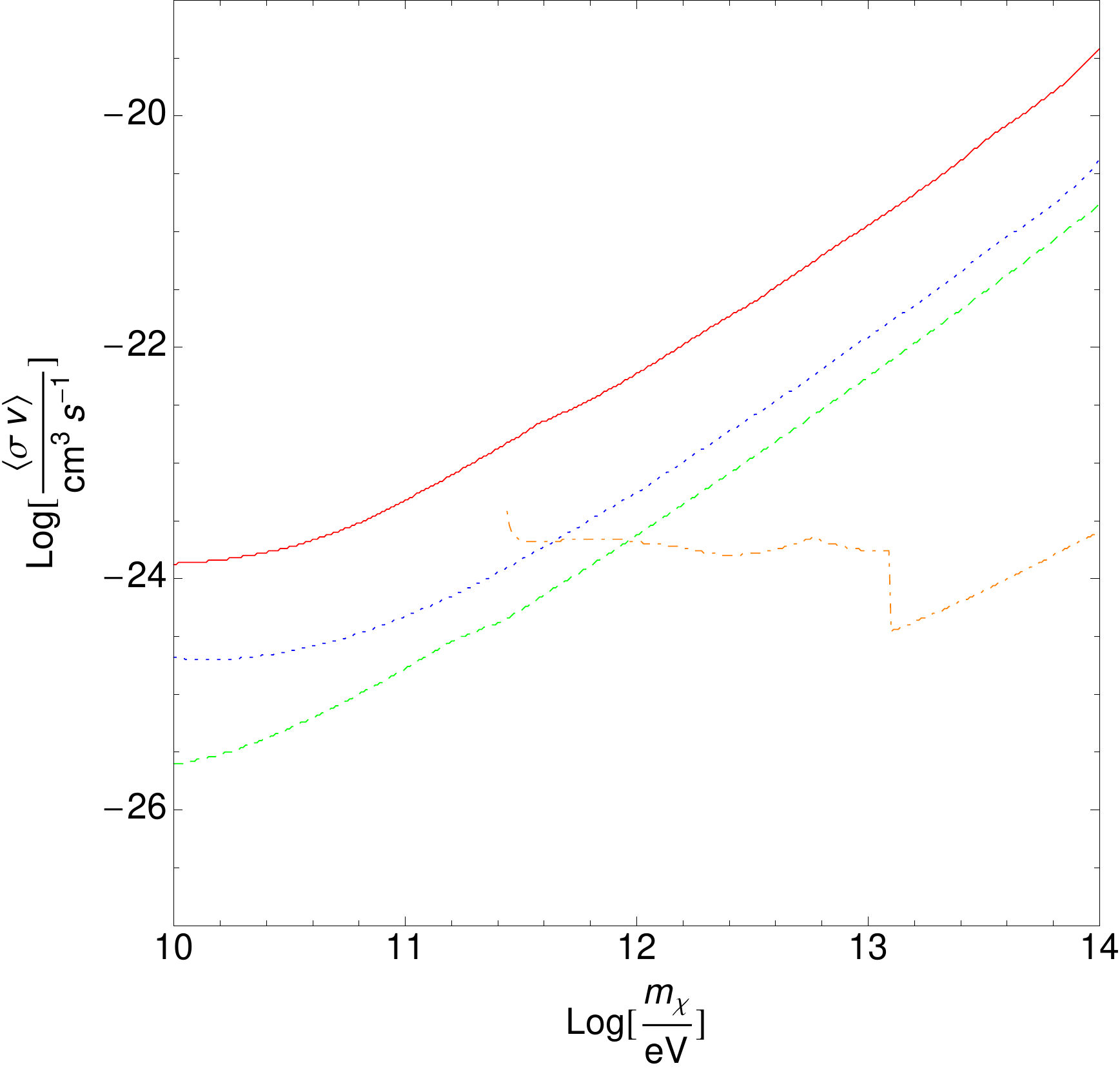}
\label{plotExclsnCntrsDNS10muGBor}
} 
\subfigure[~$\chi\chi \rightarrow \bar{q}q$; NFW profile; $B= 30 \mu$G.]{
\includegraphics[width=0.31\columnwidth]{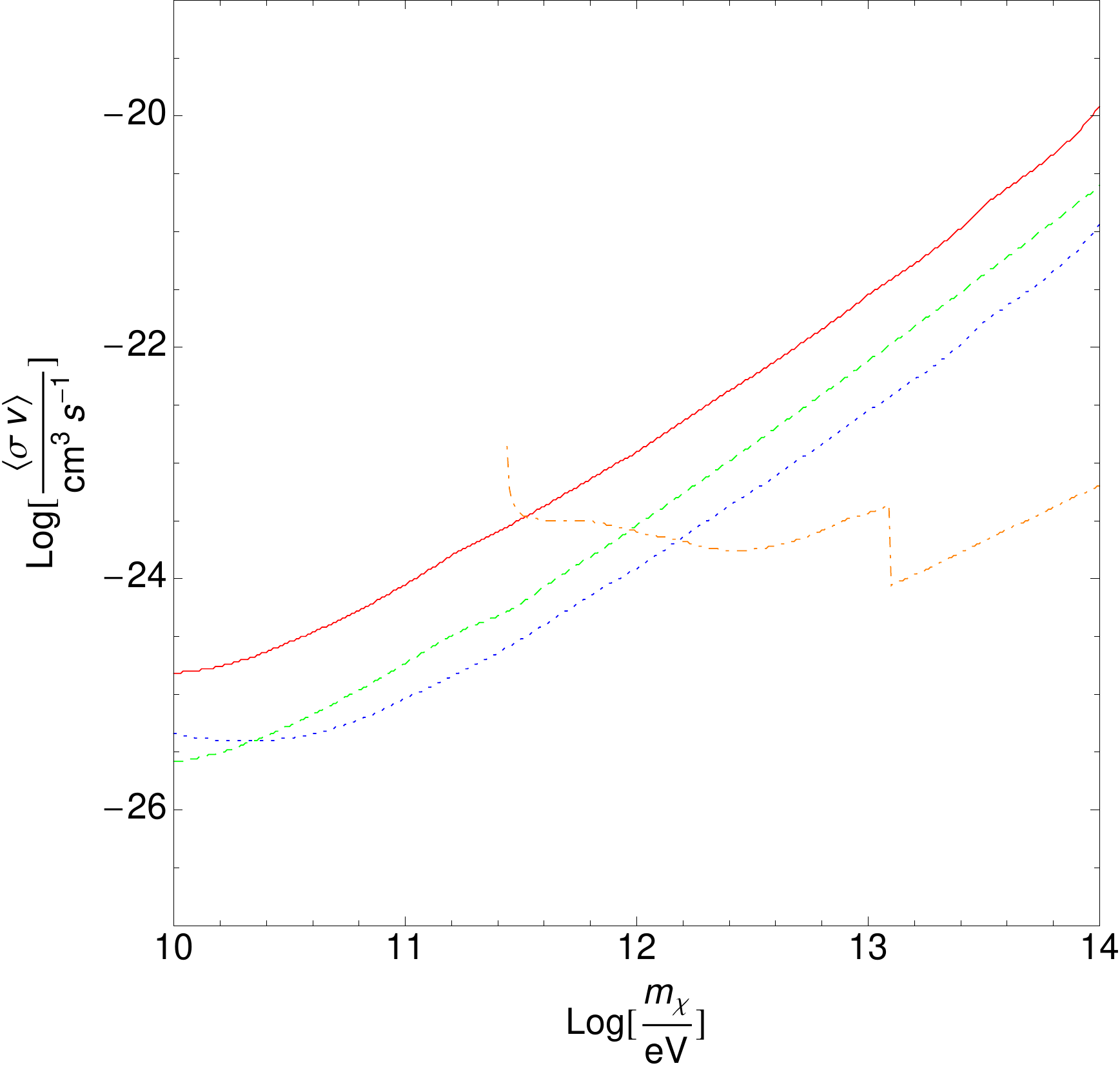}
\label{plotExclsnCntrsDNS30muGBor}
} 
\subfigure[~$\chi\chi \rightarrow \bar{q}q$; NFW profile; $B= 100 \mu$G.]{
\includegraphics[width=0.31\columnwidth]{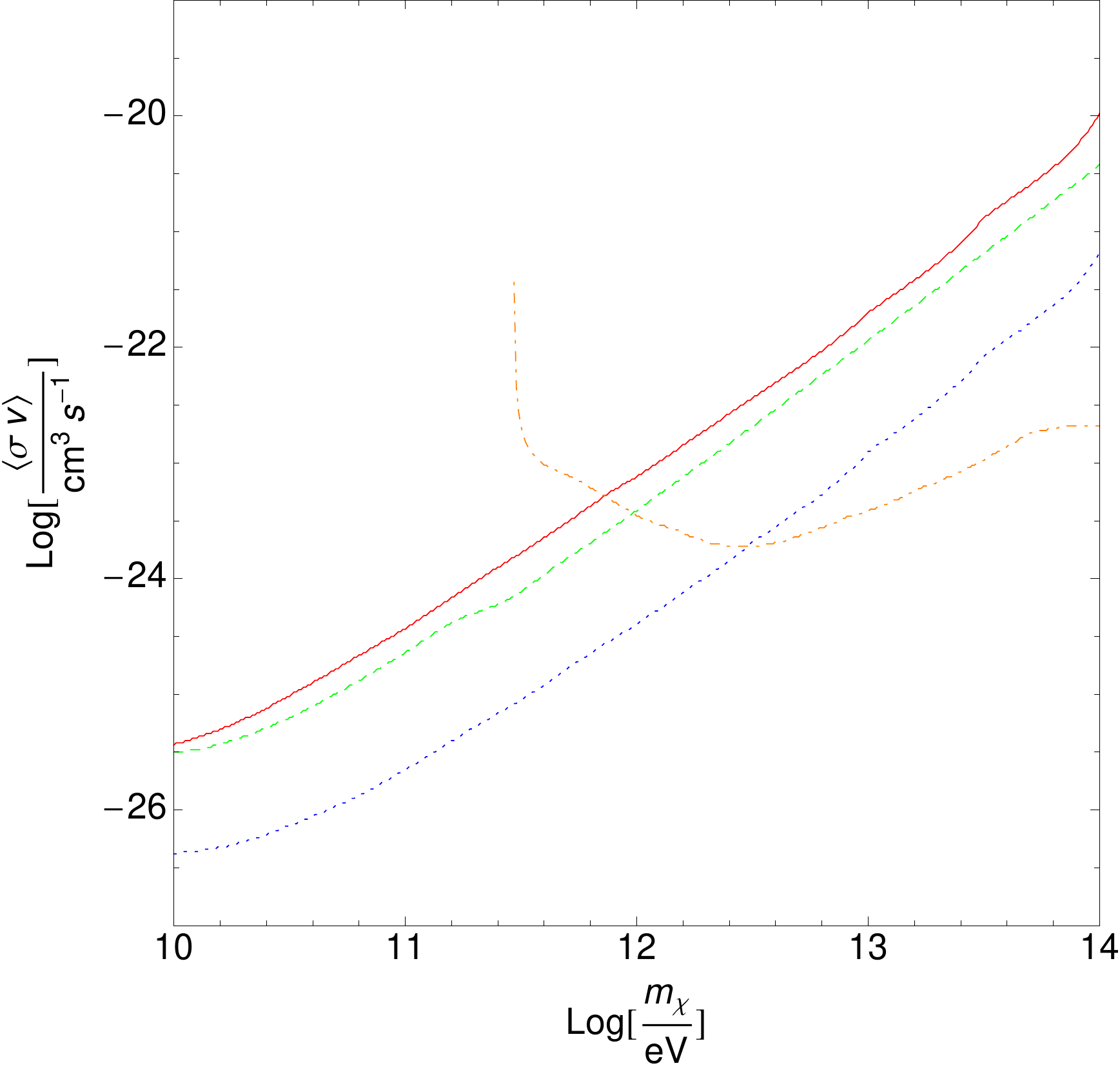}
\label{plotExclsnCntrsDNS100muGBor}
}
\caption{ Exclusion contours for the DM annihilation cross
  section assuming a $\chi\chi \rightarrow \bar{q} q$ annihilation
  channel (with a Borriello et al.~\cite{Borriello2009} $e^\pm$
  spectrum) and an NFW DM profile, obtained from DNS radio data (red, solid),
  HESS region radio data (blue, dotted), DNS region $\sim$300 MeV $\gamma$-ray data (green, dashed), and HESS region $\sim$TeV $\gamma$-ray data (orange, dot-dashed).
  Relevant $\gamma$-ray production processes are neutral meson decay, bremsstrahlung, inverse Compton emission, and internal bremsstrahlung.
  Results are displayed for three Galactic magnetic field amplitudes
  within a plausible range (a) 10 $\mu$G, (b) 30 $\mu$G and (c) 100
  $\mu$G.\label{qNFW}}
\end{figure}

\begin{figure}
\subfigure[~$\chi\chi \rightarrow e^+e^-$; NFW profile; $B= 10 \mu$G.]{
\includegraphics[width=0.31\columnwidth]{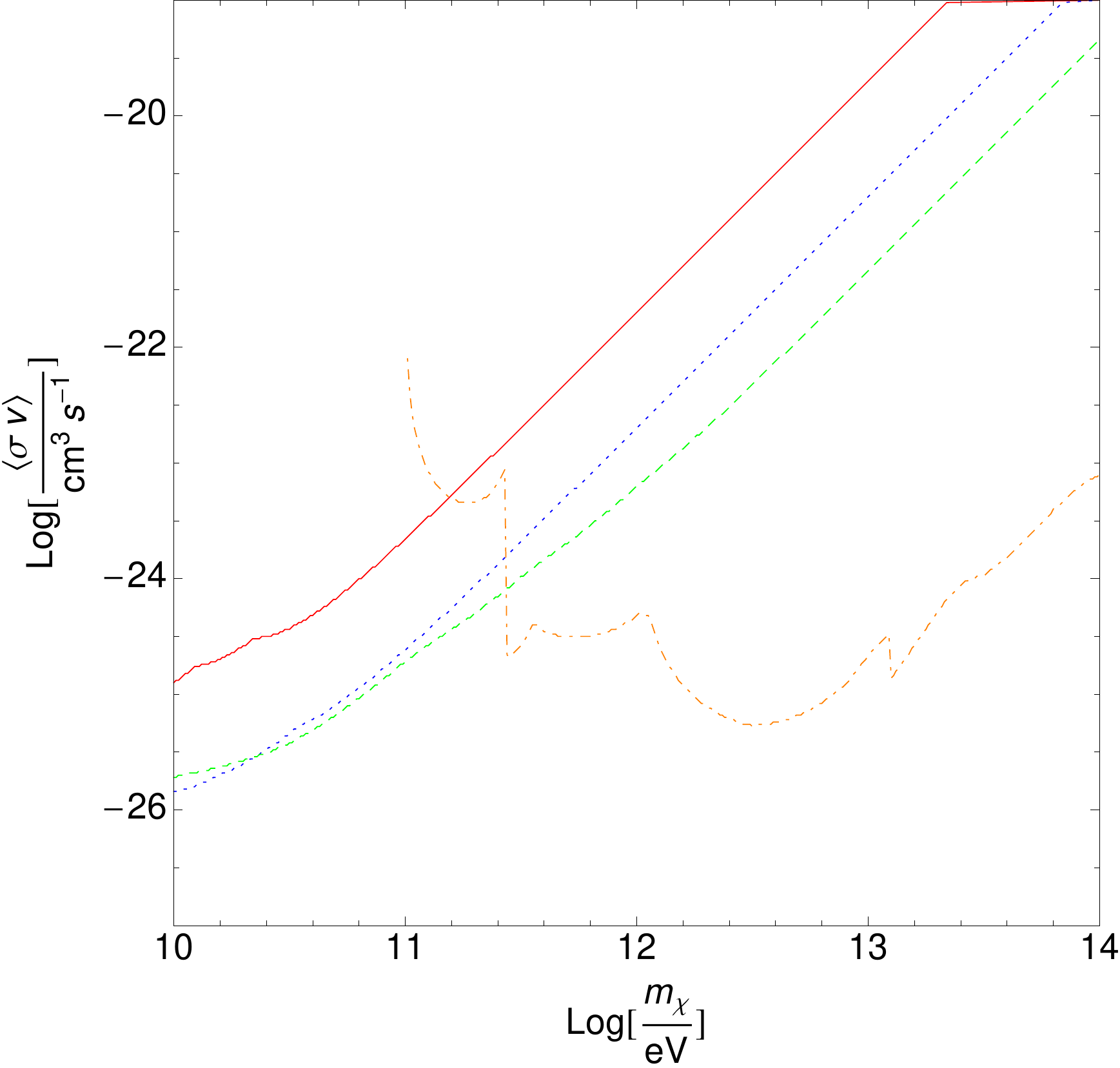}
\label{plotExclsnCntrsDNS10muGMono1}
} 
\subfigure[~$\chi\chi \rightarrow e^+e^-$; NFW profile; $B= 30 \mu$G.]{
\includegraphics[width=0.31\columnwidth]{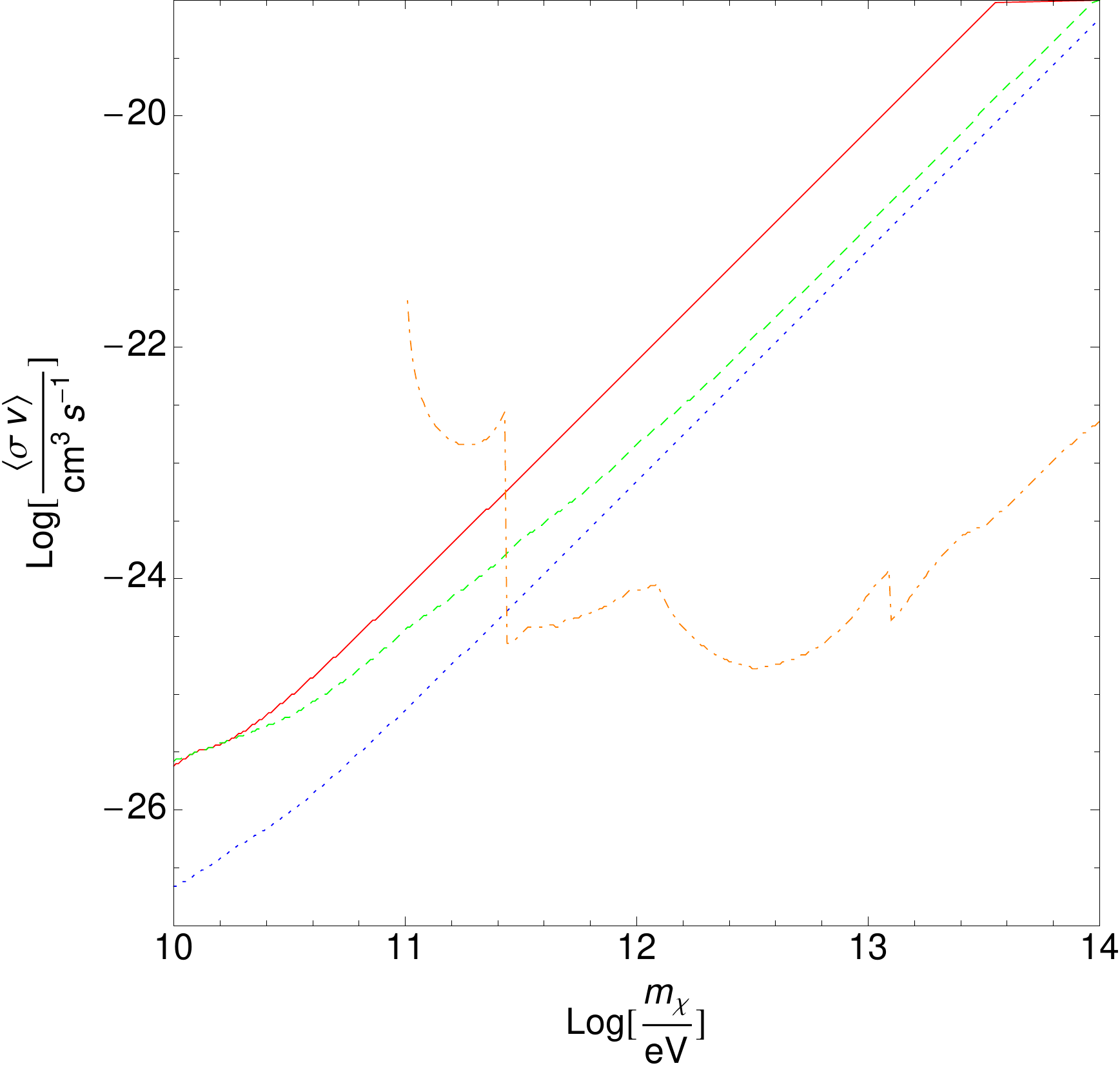}
\label{plotExclsnCntrsDNS10muGMono2}
} 
\subfigure[~$\chi\chi \rightarrow e^+e^-$; NFW profile; $B= 100 \mu$G.]{
\includegraphics[width=0.31\columnwidth]{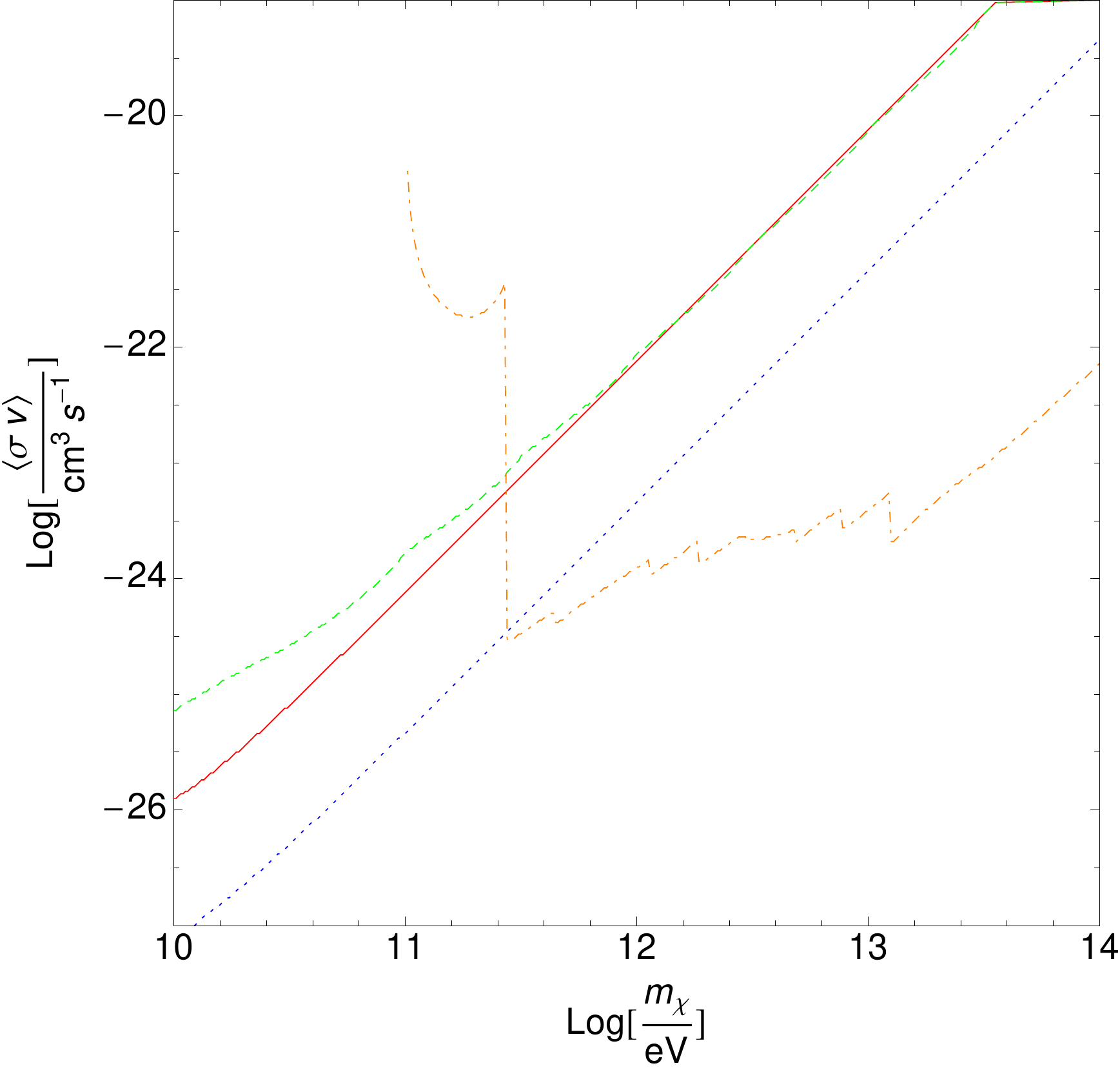}
\label{plotExclsnCntrsDNS10muGMono3}
}
\caption{Exclusion contours for the DM annihilation cross
  section, assuming a $\chi\chi \rightarrow e^+e^-$ annihilation
  channel (mono-energetic $e^\pm$ spectrum) and an NFW DM profile,
  obtained from DNS radio data (red, solid), HESS region radio data (blue, dotted),
 DNS region $\sim$300 MeV $\gamma$-ray data (green, dashed), and HESS region $\sim$TeV $\gamma$-ray data (orange, dot-dashed).
   Relevant $\gamma$-ray production processes are  bremsstrahlung, inverse Compton emission, and internal bremsstrahlung.
Results are displayed for Galactic magnetic field amplitudes
(a) 10 $\mu$G, (b) 30 $\mu$G and (c) 100 $\mu$G.\label{eNFW}}
\end{figure}

\begin{figure}
\subfigure[~$\chi\chi \rightarrow \bar{q}q$; Isothermal profile; $B = 10 \mu$G.]{
\includegraphics[width=0.31\columnwidth]{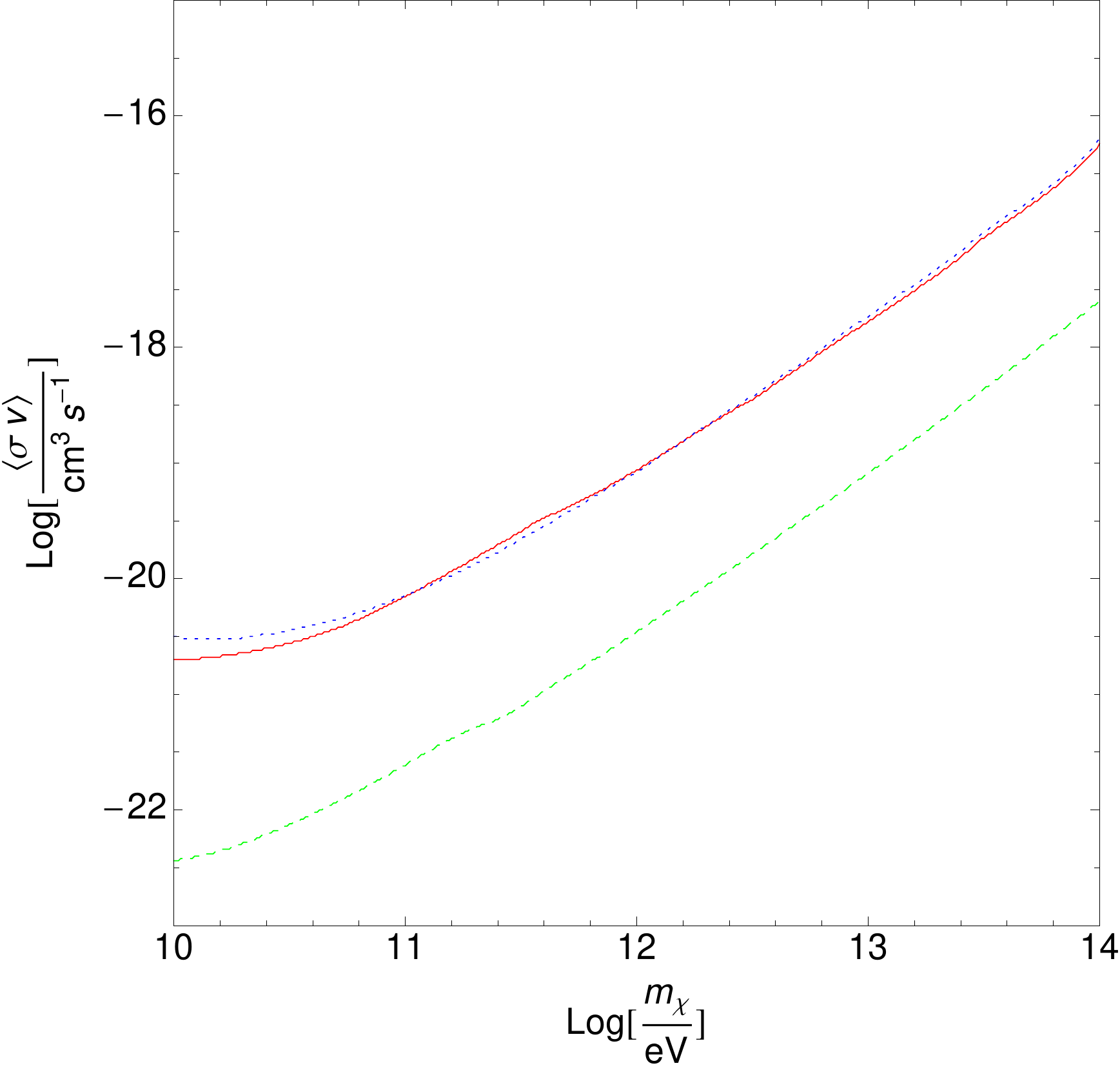}
\label{plotExclsnCntrsDNS10muGBorIso1}
} 
\subfigure[~$\chi\chi \rightarrow \bar{q}q$; Isothermal profile; $B = 30 \mu$G.]{
\includegraphics[width=0.31\columnwidth]{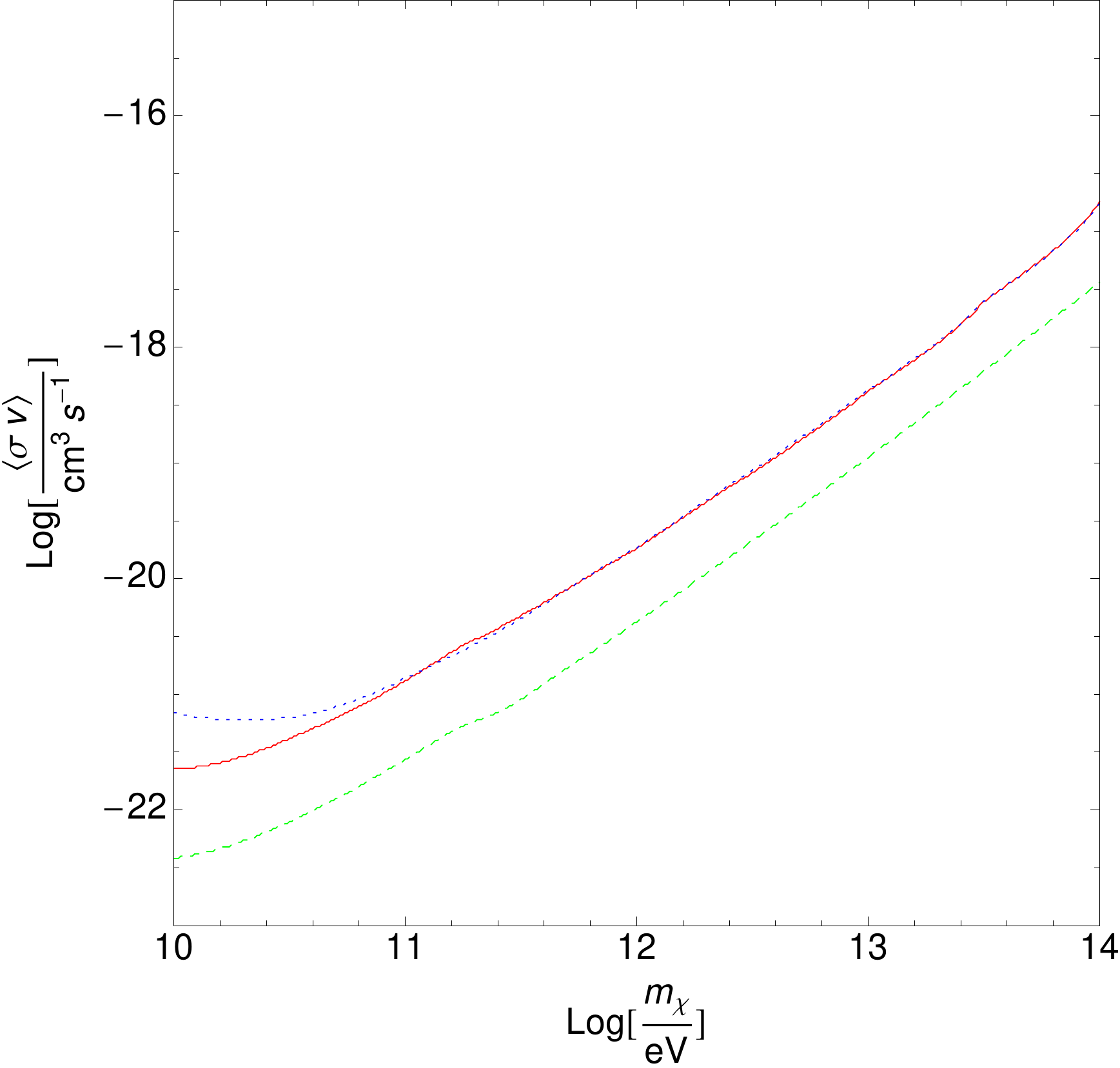}
\label{plotExclsnCntrsDNS10muGBorIso2}
} 
\subfigure[~$\chi\chi \rightarrow \bar{q}q$; Isothermal profile; $B = 100 \mu$G.]{
\includegraphics[width=0.31\columnwidth]{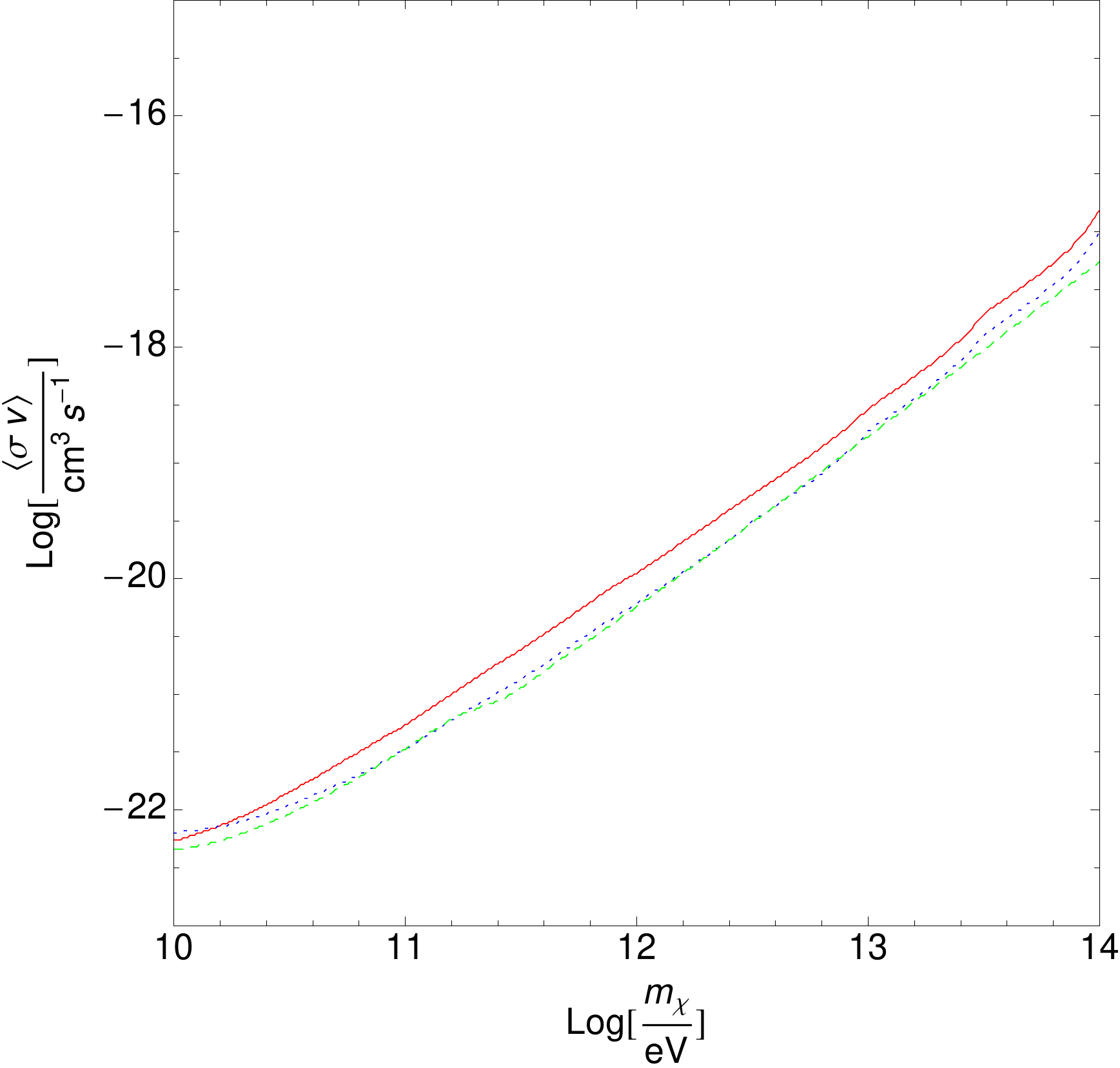}
\label{plotExclsnCntrsDNS100muGBorIso3}
}
\caption{
  Exclusion contours for the DM annihilation
  cross section assuming a $\chi\chi \rightarrow \bar{q} q$
  annihilation channel (with a Borriello et al.~\cite{Borriello2009}
  $e^\pm$ spectrum) and an Isothermal DM profile, obtained from DNS
  radio data (red, solid), HESS region radio data (blue, dotted), and $\sim$ 300 MeV $\gamma$-ray
  data (green, dashed).  [For such a flat profile the $\sim$TeV data do not offer a competitive constraint: see the text.]
Results are displayed for Galactic magnetic field amplitudes
(a) 10 $\mu$G, (b) 30 $\mu$G and (c) 100 $\mu$G.\label{qIso}}
\end{figure}

\begin{figure}
\subfigure[~$\chi\chi \rightarrow e^+e^-$; Isothermal profile; $B = 10 \mu$G.]{
\includegraphics[width=0.31\columnwidth]{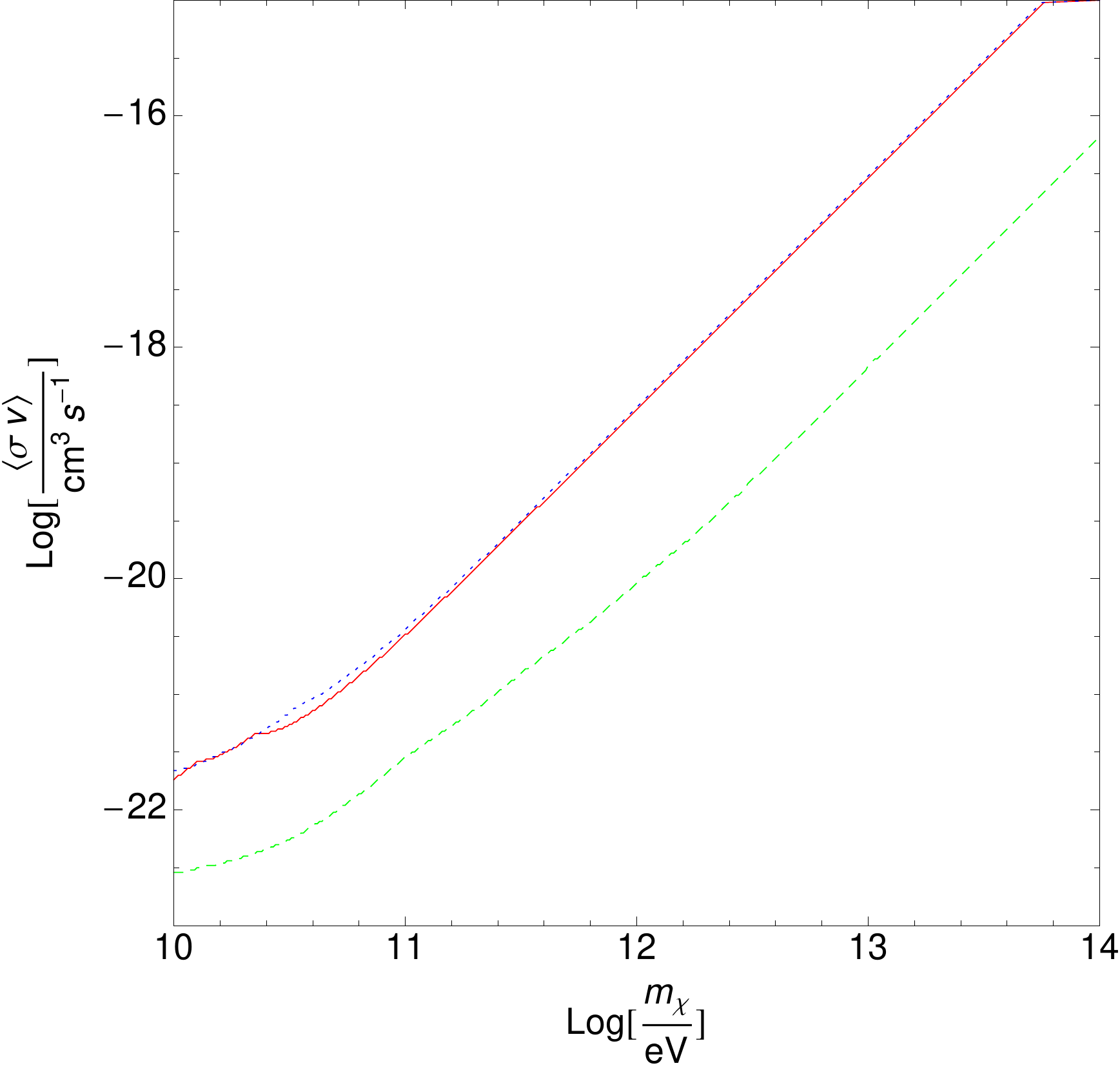}
\label{plotExclsnCntrsDNS10muGMonoIso}
} 
\subfigure[~$\chi\chi \rightarrow e^+e^-$; Isothermal profile; $B = 30 \mu$G.]{
\includegraphics[width=0.31\columnwidth]{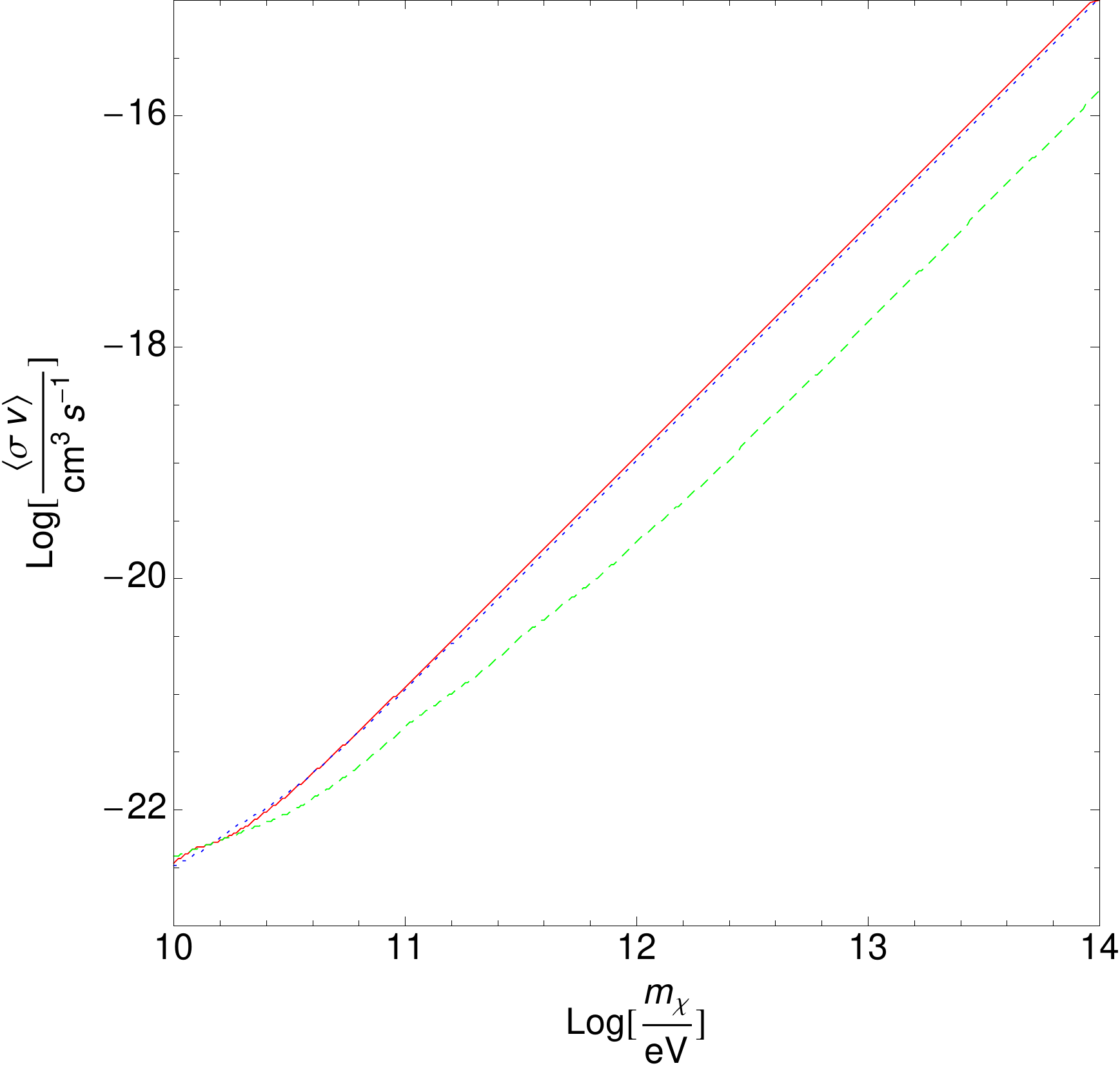}
\label{plotExclsnCntrsDNS30muGMonoIso}
} 
\subfigure[~$\chi\chi \rightarrow e^+e^-$; Isothermal profile; $B = 100 \mu$G.]{
\includegraphics[width=0.31\columnwidth]{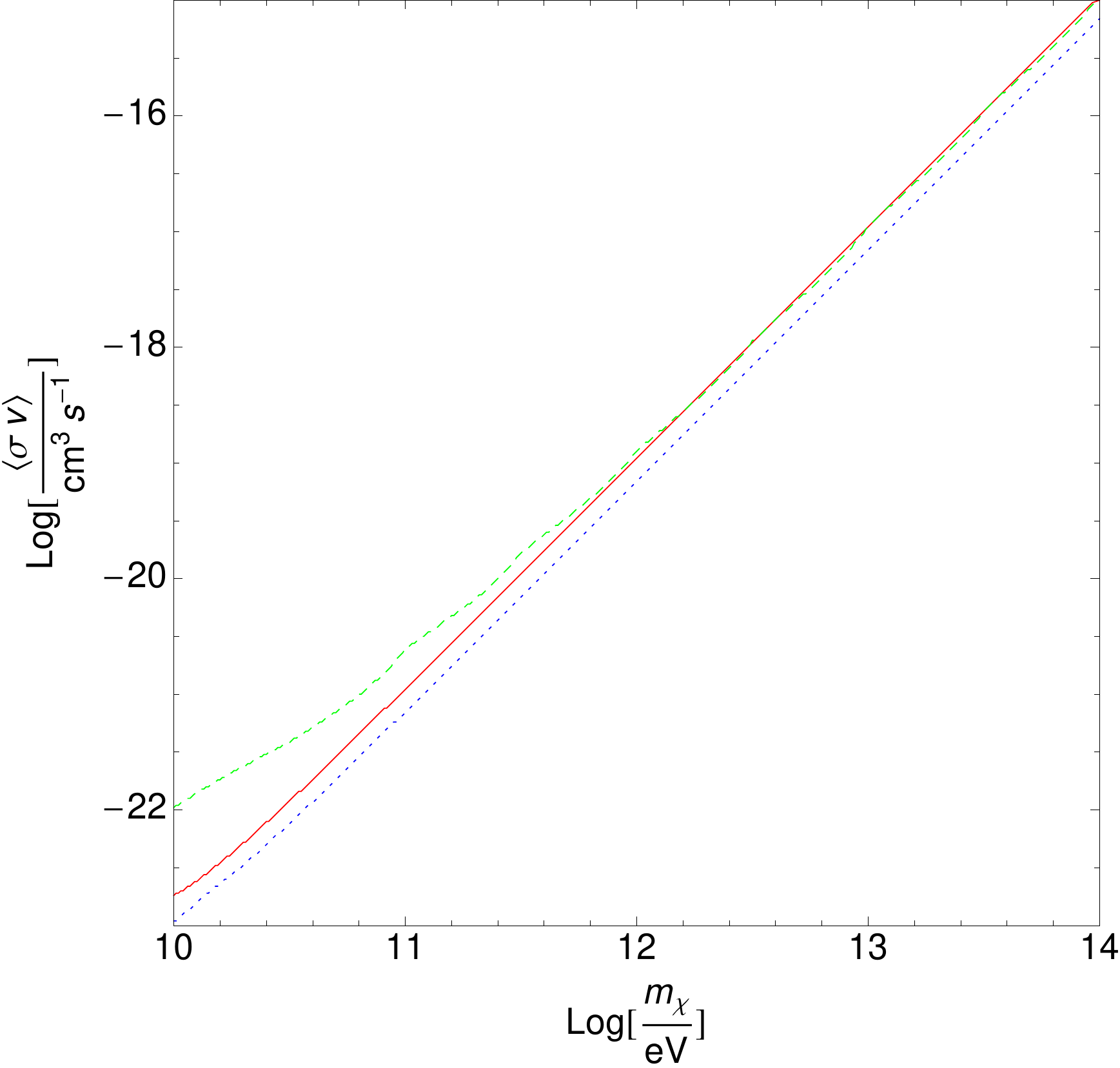}
\label{plotExclsnCntrsDNS100muGMonoIso}
}
\caption{
Exclusion contours for the DM annihilation cross
  section assuming a $\chi\chi \rightarrow e^+e^-$ annihilation
  channel (mono-energetic $e^\pm$ spectrum) and an Isothermal DM profile,
  obtained from DNS radio data (red, solid), HESS region radio data (blue, dotted),
  and  $\sim$ 300 MeV $\gamma$-ray data (green, dashed). 
  [For such a flat profile the $\sim$TeV data do not offer a competitive constraint: see the text.]
Results are displayed for Galactic magnetic field amplitudes
(a) 10 $\mu$G, (b) 30 $\mu$G and (c) 100 $\mu$G.\label{eIso}}
\end{figure}
\end{widetext}

\begin{figure}
\includegraphics[width=\columnwidth]{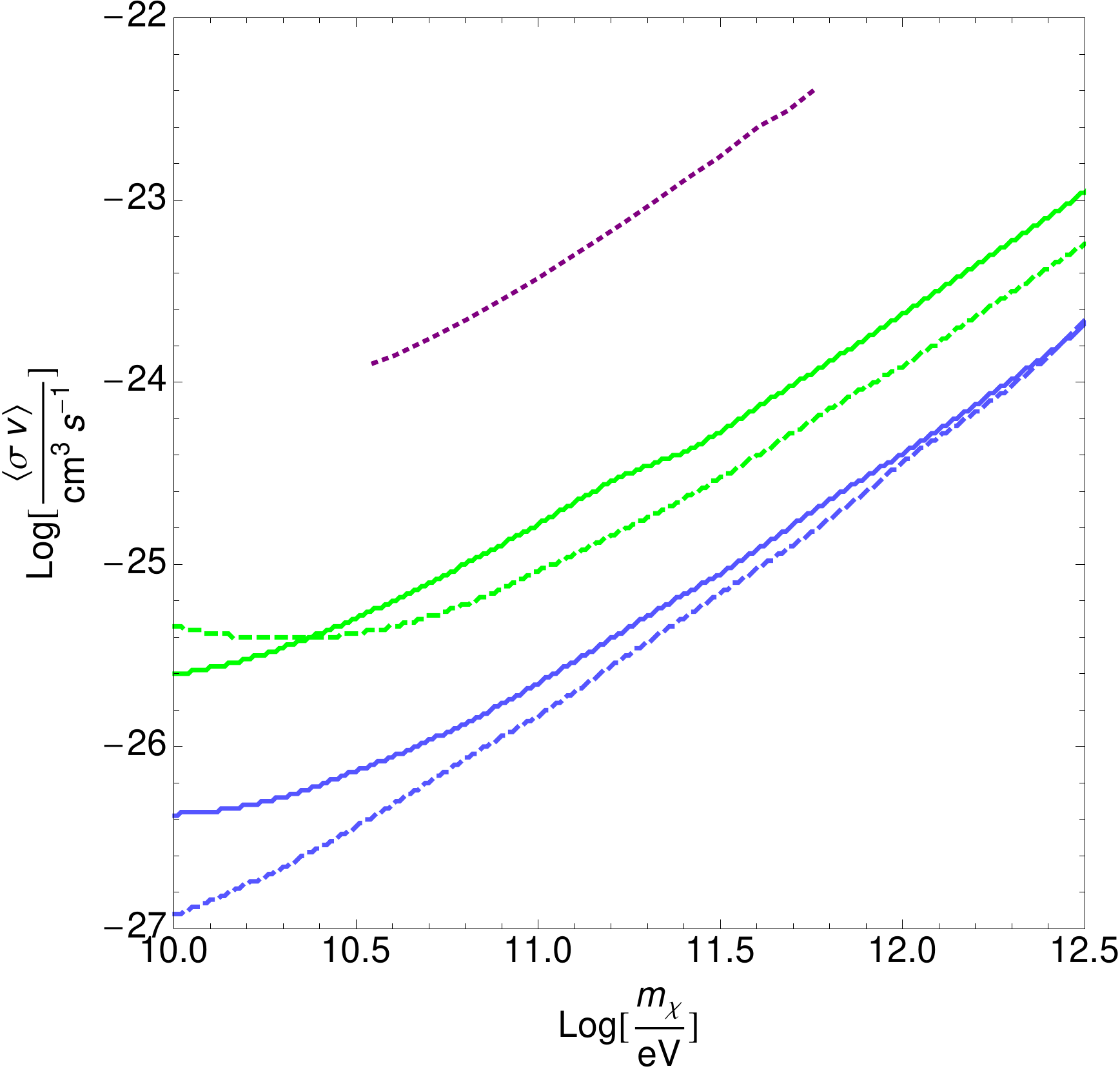}
\caption{Plot of the best constraint we find (from any modality except fot $\sim$TeV $\gamma$-rays)
  over plausible magnetic field values, assuming a $\chi\chi
  \rightarrow \bar{q} q$ annihilation channel and an NFW DM profile.
  Curves are: {\bf green} -- $\gamma$-ray constraints for (dashed) 30
  $\mu$G and (solid) 10 $\mu$G; and {\bf blue} -- radio constraints
  from the HESS region for (solid) 100 $\mu$G field and (dashed) 1 mG
  field.  Also shown for comparison is the best constraint obtained by
  Borriello et al.~\cite{Borriello2009} for this scenario, {\bf
    purple, dotted}.  }
\label{plotBestConstrntAllFields}
\end{figure}

\subsection{Comparison with previous results}

A number of previous works have considered DM annihilation corss
section constraints obtained from Galactic radio and $\gamma$-ray
data.  While our analysis is complementary to these prior studies,
distinguished by the use of different GC radio data, it is useful to
compare the strength of the limits obtained on $\langle \sigma
v\rangle$:
\begin{enumerate}
\item Of recent work, the most immediately comparable to ours -- as
  discussed above -- is that of Borriello et al.~\cite{Borriello2009}
  whose NFW profile and electron injection spectrum we adopt (for the
  $\chi\chi \rightarrow \bar{q} q$ process).
Our results are at least one order of magnitude and up to two orders
of magnitude better than those obtained by these authors. The
constraints of Borriello et al.~are obtained from consideration of
all-sky radio data at frequencies between 100 MHz and 23 GHz. These
authors, however, specifically exclude a 15$^\circ \times 15^\circ$
patch of sky centered on the GC (because their calculations are
performed in the limit that only synchrotron cooling is important in
shaping the steady state electron distribution and -- as discussed at
length above -- in the higher gas densities near the GC, this will no
longer be true given the consequent importance of ionization and
bremsstrahlung).

\item Hooper~\cite{Hooper:2008zg} has considered constraints
  obtainable from WMAP data~\cite{Spergel2007} at 22, 33, and 61
  GHz. His analysis relies on astrophysical {\it
    foreground-subtracted} data that putatively reveals a residue,
  unexplained ``haze"~\cite{Hooper2007} at all three WMAP frequencies
  on scales out to about $20^\circ$ from the GC (and absent
  elsewhere). The Hooper analysis, then, is intrinsically less
  conservative than the radio analysis presented here where we do not
  attempt such subtraction (except for discrete sources in the field
  which can be unambiguously identified on angular size
  grounds). 
While Hooper's constraints are particularly strong, some of the
astrophysical assumptions adopted may be too optimistic.  In particular
he assumes a constant 10 $\mu$G field and constant 5 eV cm$^{-3}$
interstellar radiation field energy density over the 30$^\circ$
angular region he investigates, the former being rather stronger and
the latter rather weaker (at least towards the GC) than assumed by
other authors~\cite{Borriello2009,Grajek:2008jb}.
Grajek et al.~\cite{Grajek:2008jb} consider the same
(foreground-subtracted) WMAP data investigated by Hooper.  Assuming a
rather weaker (and spatially-varying) magnetic field and a rather
smaller ratio $U_B/(U_B + U_\textrm{rad}) \sim 0.1$, 
these authors find correspondingly weaker constraints on the
annihilation cross section, which we surpass.  
(Note that the $W^+W^-$ annihilation channel considered by Hooper and
Grajek results in a very similar electron injection spectrum to the
$\bar{q}q$ channel we consider.)

\item Finally, Bertone {\it et al.}\cite{Bertone2009} and Regis and
  Ullio~\cite{Regis2009} both also consider constraints arising from
  multi-wavelength observations of the GC, including radio and
  $\gamma$-rays. These papers tend to focus on observations on much
  smaller angular scales around the super massive black at the
  Galactic dynamical center than investigated here.
  While some of the constraints (and projected constraints) derived are
  quite strong, on these small scales it is necessary to assume a particular model for the
  evolution of the magnetic field intensity (and the matter density controlling bremsstrahlung and ionization cooling).  
  Such constraints are also particularly sensitive to the DM density at
  extremely small radii, which is highly uncertain.

\end{enumerate}

We now compare with constraints derived using other techniques.   
For hadronic annihilation modes, constraints arising from the
antiproton observations have been shown to be of comparable
sensitivity to existing Galactic centre gamma ray and radio limits,
e.g.  Refs.~\cite{Pato:2009fn,Catena:2009tm}, for some models.
For the monoenergetic $e^+e^-$ annihilation channel, pure IB
constraints were derived in Ref.~\cite{bell_jacques}.  For the NFW
profile, the IB constraints are very strong for $10^3 \alt m_\chi \alt
10^4$ (these arise from HESS gamma ray data near the GC) whereas our
new radio/gamma ray bounds are better at lower mass.  (For an
isothermal profile, the IB constraints arising from the HESS data
would be considerable weaker.)
Note that for masses above $\sim 10$ TeV, bounds on the total
annihilation cross section to all final states, derived using neutrino
data, become most constraining~\cite{BBM,YHBA,KS}.  For higher masses,
$m_\chi \agt 10^2$, TeV the unitarity bound on the total cross section
becomes the most restrictive constraint~\cite{Hui}.

Turning now to {\it non}-Galactic based techniques, a number of
authors have recently considered constraints arising from annihilation
at high redshift.  For example,
Refs.~\cite{Belikov:2009qx,Galli:2009zc,Huetsi:2009ex,Cirelli:2009bb,Slatyer:2009yq}
consider the effect of energy deposited by DM annihilation during the
the reionization and recombination epochs.
(Note that these techniques are complementary to those we consider
here, being subject to quite different assumptions and systematics.)
For an NFW profile and an $e^+e^-$ annihilation channel the
reionization limits are broadly comparable to the constraints we have
derived here, with the boost factor required to be $\alt 10$ at
$m_\chi\sim 100$ GeV.

\section{Summary and conclusions}\label{Conclusion}

We have considered DM annihilation in the Galactic center, for
scenarios in which annihilation leads to the production of
relativistic electrons and positrons.  Using signals arising from
electron energy loss processes, namely synchrotron emission,
bremsstrahlung and inverse Compton scattering (and, where appropriate,
those due to neutral meson decay and internal bremsstrahlung), we have
derived robust constraints on the velocity-averaged DM self
annihilation cross section $\langle \sigma v\rangle$.
The processes considered were $\chi\chi \rightarrow e^+e^-$ which
produces monoenergetic $e^\pm$, and $\chi\chi \rightarrow \bar{q} q$,
in which a spectrum of $e^\pm$ are generated via the charged pion
decay chain.  (The constraints for the later channel are expected to be
very similar to those for other interesting final states, such as
$W^+W^-$ and $ZZ$.)

We have demonstrated that a combination of radio and gamma ray bounds
is relatively insensitive to the assumed magnetic field amplitude
within a plausible range, with the constraint on $\langle \sigma v
\rangle$ varying by a factor of $< 10$ as $B$ is varied from 10 -
100 $\mu$G.  (Taken alone, the radio and gamma rays bounds are
individually quite sensitive to the assumed magnetic field amplitude.)
Our analysis is distinct from previous work in this area by the use of
a new synthesis of Galactic center radio data assembled in
Ref.~\cite{Crocker2009}.

Our constraints on velocity-averaged DM annihilation cross sections
are conservative as we {\it (i)}
do not remove (known) astrophysical contributions to the radio
  emission and {\it (ii)}
%
do not include a contribution to the total radio emission from
  DM annihilation electrons and positrons along the line-of-sight but
  out of the GC.




Despite these conservative assumptions, our results are at least one
order of magnitude and up to two orders of magnitude better than the
most directly comparable previous limit (obtained by Borriello et
al. for the same DM distributions and electron injection spectrum, but
considering all-sky radio fluxes).  Moreover, our constraints rule out
a sizable portion of the parameter space that has been invoked to
explain the various positron anomalies.  (See,
e.g. Refs.~\cite{Cirelli:2009bb,Meade:2009iu} for recent
determinations of the PAMELA/Fermi perferred regions.)

For an NFW profile and a $\chi\chi \rightarrow \bar{q}q$ annihilation
process, we find the allowed boost factor is $\alt 1$ at 10 GeV and
$\alt 100$ at 1 TeV.  For $\chi\chi \rightarrow e^+e^-$, the allowed
boost factor is $< 1$ at 10 GeV and $< 1000$ at 1 TeV.
Note that these constraints apply to boost factors generated via
whatever mechanism, be it a Breit-Wigner resonance, the Sommerfeld
effect, or an enhancement due to DM substructure~\footnote{An
  interesting exception is the case where we can not reliably
  extrapolate between the local DM annihilation rate and that in the
  Galactic center, e.g., because of a variation in the dispersion
  velocity~\cite{Cholis:2009va} or a local clump of dark matter.}.

\vspace{10mm}

\section*{Acknowledgments}
The authors are indebted to T. Jacques, A. Mazumdar, N. Sahu, and F. Wang for
invaluable discussions on various aspects of dark matter models and
the indirect detection of dark matter signals.  This research was
funded in part by the Australian Research Council under Project ID
DP0877916 and DP0988343.

\appendix
\section{Electron transport}
\label{section_thickTarget}
Much research concerning cosmic ray electrons adopts the approach of taking measured or inferred values of the ``diffusion coefficient'' (for cosmic ray electrons at a particular energy), assuming some energy scaling for this quantity, and then deriving characteristic distances over which electrons might diffuse over their cooling times. There are many uncertainties attendant upon this statistical approach, particularly for the Galactic center environment where evidence points to ambient magnetic fields stronger and more turbulent than found typically in the disk. We prefer to take the approach of referring to direct modeling (by solving the Lorentz force equation) of individual particle trajectories for the GC environment.
Here we rely on the work of Wommer et al.~\cite{Wommer2008} who consider {\it proton} propagation in this environment.
  From the work of these authors we infer that $\cal{O}$(TeV) energy protons might travel $\lesssim 0.2^\circ$ or 30 pc (at the GC distance) over their radiative lifetimes for an environment with a 10 $\mu$G ambient field and $n_H = 100$ cm$^{-3}$. In this environment the TeV proton cooling timescale (due to ionization and $p-p$ collisions) is  $\sim 4 \times 10^5$ years. In contrast, for GeV scale electrons (synchrotron-radiating at about GHz frequencies), the  cooling timescale is {\it less} than this at $\lesssim 2 \times 10^5$ years for the weakest magnetic field (10 $\mu$G) and most tenuous gas environment ($n_H \simeq 3$ cm$^{-3}$) we consider. Given that both TeV protons and GeV electrons are ultra-relativistic and that diffusion distance will decline towards lower particle energies, we infer from Ref.~\cite{Wommer2008} that GeV electrons will travel well less than $0.2^\circ$ or 30 pc in their radiative lifetimes. Given this angular scale is well smaller than both the DNS and HESS fields we conclude that it is safe to conduct our calculations in the thick target limit.



\end{document}